\documentclass{aa}  
\usepackage{graphicx}
\usepackage{txfonts}
\begin{document} 

   \title{Measuring the gas reservoirs in $10^{8}<$ M$_\star<10^{11}$~M$_\odot$ galaxies at $1\leq z\leq3$}

   \author{Rosa M. M\'erida
          \inst{1}\inst{2}
          \and
          Carlos G\'omez-Guijarro\inst{3}\and
          Pablo G. P\'erez-Gonz\'alez \inst{1}\and Patricia S\'anchez-Bl\'azquez \inst{4}\inst{5} \and David Elbaz \inst{3} \and Maximilien Franco\inst{6} \and Lucas Leroy \inst{3} \and Georgios E. Magdis \inst{7}\inst{8}\inst{9}
          \and Benjamin Magnelli\inst{3} \and Mengyuan Xiao\inst{10}}

   \institute{Centro de Astrobiolog\'ia (CAB), CSIC-INTA, Ctra. de Ajalvir km 4, Torrej\'on de Ardoz, E-28850, Madrid, Spain; \email{rmerida@cab.inta-csic.es} \and
   Departamento de F\'isica Te\'orica, Universidad Aut\'onoma de Madrid, E-28049, Cantoblanco (Madrid), Spain \and Universit{\'e} Paris-Saclay, Universit{\'e} Paris Cit{\'e}, CEA, CNRS, AIM, 91191, Gif-sur-Yvette, France \and Departamento de Física de la Tierra y Astrof\'isica, Fac. CC. F\'isicas, Universidad Complutense de Madrid, Plaza de las Ciencias 1, Madrid, E-28040, Spain \and IPARCOS (Instituto de F\'isica de Partículas y del Cosmos),
Facultad de Ciencias F\'isicas, Ciudad Universitaria, Plaza de las Ciencias, 1, Madrid, E-28040, Spain \and The University of Texas at Austin, 2515 Speedway Blvd Stop C1400, Austin, TX 78712, USA \and Cosmic Dawn Center (DAWN), Jagtvej 128, DK2200 Copenhagen N, Denmark \and DTU-Space, Technical University of Denmark, Elektrovej 327, DK2800 Kgs. Lyngby, Denmark \and Niels Bohr Institute, University of Copenhagen, Jagtvej 128, DK-2200 Copenhagen N, Denmark \and Department of Astronomy, University of Geneva, Chemin Pegasi 51, 1290 Versoix, Switzerland\\}
    
   \date{Received September 15, 1996; accepted March 16, 1997}

 
  \abstract
   {Understanding the gas content in galaxies, its consumption and replenishment, remains pivotal in our comprehension of the evolution of the Universe. Numerous studies have addressed this, utilizing various observational tools and analytical methods. These include examining low-transition $^{12}$CO millimeter rotational lines and exploring the far-infrared and the (sub-)millimeter emission of galaxies. With the capabilities of present-day facilities, much of this research has been centered on relatively bright galaxies.}
   {We aim at exploring the gas reservoirs of a more general type of galaxy population at $1.0<z<3.0$, not restricted to bright (sub-)millimeter objects. We strive to obtain a measurement that will help to constrain our knowledge of the gas content at $10^{10-11}$~M$_\odot$, and upper limits at $\sim10^{8-10}$~M$_\odot$.}
   {We stack ALMA 1.1~mm data to measure the gas content of a mass-complete sample of galaxies down to $\sim10^{8.6}$~M$_\odot$ at $z=1$ ($\sim10^{9.2}$~M$_\odot$ at $z=3$), extracted from the HST/CANDELS sample in GOODS-S. The selected sample is composed of 5,530 on average blue ($<b-i>\sim0.12$ mag, $<i-H>\sim0.81$ mag), star-forming main sequence objects ($\Delta$MS=log SFR - log SFR$_{\mathrm{MS}}\sim-0.03$).}
   {At $10^{10-11}$~M$_\odot$, our gas fractions (f$_{\mathrm{gas}}$=M$_{\mathrm{gas}}$/(M$_{\mathrm{gas}}$+M$_\star$)), ranging from 0.32 to 0.48 at these redshifts, agree well with other studies based on mass-complete samples down to $10^{10}$~M$_\odot$, and are lower than expected according to other works more biased to individual detections. At $10^{9-10}$~M$_\odot$, we obtain 3$\sigma$ upper limits for f$_{\mathrm{gas}}$ which range from 0.69 to 0.77, and at $10^{8-9}$~M$_\odot$ these upper limits rise to $\sim0.97$. The upper limits at $10^{9-10}$~M$_\odot$ are on the level of the extrapolations of scaling relations based on mass-complete samples and below those based on individual detections. As such, it suggests that the gas content of low-mass galaxies is at most what is extrapolated from literature scaling relations based on mass-complete samples down to $10^{10}$~M$_\odot$. Overall, the comparison of our results with previous literature reflects how the inclusion of bluer, less obscured, and more MS-like objects progressively pushes the level of gas to lower values.}
   {}

   

   \maketitle
%

\section{Introduction}
\label{sec:intro}
Cold molecular gas is the material that fuels the galaxy machinery that works to form stars. Knowing the amount of gas available in galaxies, how efficiently it is converted into stars, as well as how it is replenished is crucial to understanding their evolutionary pathways. The cosmic history of the gas mass density resembles that of the star formation rate density (\citealt{Decarli2019}, \citealt{Riechers2019}, \citealt{Magnelli2020}, \citealt{Walter2020}), peaking at $z \sim 2$ and steadily decreasing until now. The gas mass (M$_{\mathrm{gas}}$) content in galaxies at a fixed stellar mass (M$_\star$) increases with redshift at least at $0 < z < 3$. At a fixed redshift, the gas fraction (f$_{\mathrm{gas}}$ = M$_{\mathrm{gas}}$/(M$_\star$+M$_{\mathrm{gas}}$)) decreases with M$_\star$ (\citealt{Genzel2010}, \citealt{Bethermin2015}, \citealt{MorokumaMatsui2015}, \citealt{DessaugesZavadsky2020}, \citealt{Tacconi2020}, \citealt{Magnelli2020}, \citealt{Wang2022}). 

The relation between M$_{\mathrm{gas}}$ and M$_\star$ at different redshifts has been quantified by a variety of studies (e.g., \citealt{Scoville2016}, \citealt{Tacconi2018}, \citealt{Liu2019b}, \citealt{Tacconi2020}, \citealt{Kokorev2021}, \citealt{Wang2022}), covering $0 < z < 6$. It is typically parameterized according to cosmic time or redshift, and the distance from the galaxies to the main sequence (MS) of star-forming galaxies (SFGs). The term MS refers to the tight correlation that exists between the SFR and the M$_\star$ (e.g., \citealt{Noeske2007}, \citealt{Elbaz2011}, \citealt{Whitaker2012}, \citealt{Speagle2014}, \citealt{Tomczak2016}, \citealt{Santini2017}, \citealt{Pearson2018}, \citealt{Barro2019}), which is seen to be present at least at $0 < z < 6$.

The cold molecular gas can be studied directly using rotational lines of molecular hydrogen, H$_2$. However, the transition probabilities are very small, line emission is weak, and transitions are sufficiently excited only in radiation or shock-warmed molecular gas-like photodissociation regions and outflows (\citealt{Parmar1991}, \citealt{Richter1995}). Common alternatives to study the gas content in distant galaxies include the use of the low-transition $^{12}$CO millimeter rotational lines and dust continuum measurements.

For the first approach, it is typically assumed that the CO lower rotational lines are optically thick and the CO line luminosity is proportional to the total molecular gas mass (M$_{\mathrm{H_2}}$), using an empirical conversion factor (\citealt{Dickman1986}, \citealt{Solomon1987}, \citealt{Bolatto2013}). 

For the second approach, the M$_{\mathrm{gas}}$ can be derived based on the dust content, converting the dust mass (M$_\textrm{dust}$) obtained by fitting the infrared (IR) spectral energy distribution (SED) \citep{DraineLi2007} to M$_{\mathrm{gas}}$, for which it is typically assumed a metallicity-dependent gas-to-dust ratio ($\delta_{\mathrm{GDR}}$; e.g. \citealt{Magdis2012}, \citealt{Genzel2015}). One can also use the photometry measured in the Rayleigh–Jeans (RJ) tail of the SED (e.g. \citealt{Scoville2016}, \citealt{Hughes2017}). The \citet{Scoville2016} method (S16 hereafter) works similarly to the previous one, assuming a constant $\delta_{\mathrm{GDR}}$ with a mass-weighted dust temperature (T$_{\mathrm{dust}}$) of 25~K. These approaches assume that zero-point calibrations based on $z = 0$ measurements are also valid at higher redshifts.

These methods have been previously used in other works to study the gas content in the local and distant universe (e.g. \citealt{Saintonge2017}, \citealt{Decarli2019}, \citealt{Freundlich2019}, \citealt{Sanders2023} derived M$_{\mathrm{gas}}$ using the $^{12}$CO rotational lines; \citealt{Schinnerer2016}, \citealt{Wiklind2019}, \citealt{Kokorev2021} used the dust emission; and \citealt{Liu2019b}, \citealt{Aravena2020}, and \citealt{Birkin2021} used both methods).
However, despite the increasing number of studies in the field, most of the efforts so far focus on individual (sub-)millimeter detections of massive objects ($>10^{10-11}$~M$_\odot$). For instance, the \citealt{Schinnerer2016} sample is made up of ALMA detections at 240~GHz, with M$_\star>10^{10.7}$~M$_\odot$ at $z \sim 3.2$. In \citet{Freundlich2019}, they include CO emitters with 10$^{10-11.8}$~M$_\odot$ within $0.5 < z < 3$. The \citealt{Liu2019b} sample contains galaxies at $0.3 < z < 6$ that show high-confidence ALMA detections, with median M$_\star = 10^{10.7}$M$_\odot$.

Alternatively, other studies sought to extend this analysis to fainter galaxies and improve the completeness of the data-sets by stacking the emission of similar sources, not imposing a flux criterion on the (sub-)millimeter emission of the sources. In \citet{Tacconi2020}, part of their sample is based on this strategy, made up of stacks of $Herschel$ far infrared (FIR) spectra. Their data-set also includes individual CO emitters though. In \citet{Magnelli2020}, they measure the cosmic density of dust and gas by stacking $H$-band selected galaxies above a certain M$_\star$. In \citet{Garratt2021}, they study the evolution of the H$_2$ mass density back to $z \approx 2.5$ measuring the average observed 850~$\mu$m flux density of near-infrared selected galaxies. In \citet{Wang2022}, they employ stacking to derive the mean mass and extent of the molecular gas of a mass-complete sample down to 10$^{10}$~M$_\odot$. In the latter study, they obtain M$_{\mathrm{H}_2}$ which are generally lower than previous estimates, based on individual detections.

In this work, we use the emission at 1.1~mm measured with observations obtained by the Atacama Large Millimeter/submillimeter Array (ALMA) to infer the content of gas present in a mass-complete sample of galaxies at $1.0 < z < 3.0$, analyzing stacked ALMA images for subsamples in different redshift ranges and M$_\star$ bins. Taking advantage of the galaxy catalog provided by the Cosmic Assembly Near-infrared Deep Extragalactic Legacy Survey (CANDELS; \citealt{grogin2011}, \citealt{koekemoer2011}) in The Great Observatories Origins Deep Survey (GOODS; \citealt{Dickinson2003}, \citealt{Giavalisco2004}), specifically in GOODS-S (\citealt{Guo2013}, G13 hereafter), we probe the $10^{10-11}$~M$_\odot$ stellar mass regime with a complete sample whose 80\% completeness level reaches down to 10$^{8.6}$~M$_\odot$ at $z = 1$ (10$^{9.2}$~M$_\odot$ at $z = 3.0$) \citep{Barro2019}. Our analysis aims at removing potential biases at the high-mass end when based on detections of individual galaxies. We aspire to check whether faint sources in ALMA follow the same scaling relations derived from brighter sources or, on the contrary, present a distinct molecular gas content than that prescribed for their stellar masses. Moreover, this sample gives us the chance to explore the gas reservoirs of less massive galaxies, $\sim10^{9-10}$~M$_\odot$, for which previous scaling relations are still not well calibrated.

The structure of the paper is as follows. In Section \ref{sec:sample} we present the data and sample selection. We then describe the physical properties of the sample and compare them with other catalogs in Section \ref{sec:properties}. In Section \ref{sec:stacking} we present our stacking and flux measurement methodology applied to the ALMA data. In Sections \ref{sec:gas_prop} and \ref{sec:discussion}, we present and discuss our results regarding the gas reservoirs of our sample, comparing them with previous scaling relations. The conclusions are summarized in Section \ref{sec:conclusions}.

Throughout the paper we assume a flat cosmology with $\Omega_\mathrm{M}=0.3$, $\Omega_{\lambda}=0.7$ and a Hubble constant H$_0=70$ km s$^{-1}$ Mpc$^{-1}$. We use AB magnitudes \citep{Oke1983}. All M$_\star$ and SFR estimations refer to a \citet{Chabrier2003} initial mass function (IMF).

\section{Data and sample} \label{sec:sample}

\subsection{Data}

We base this work on the images provided by the GOODS-ALMA 1.1~mm galaxy survey (\citealt{Franco2018}, \citealt{carlos2022a}) in GOODS-S, carried out in ALMA Band 6. This survey extends over a continuous area of 72.42~arcmin$^2$ (primary beam response level
$\geq20$\%) with a homogeneous average sensitivity. It is the result of two different array configurations. Cycle 3 observations (program 2015.1.00543.S; PI: D. Elbaz) were based on a more extended array configuration that provided a high angular resolution data-set (\citealt{Franco2018}). Cycle 5 observations (program 2017.1.00755.S; PI: D. Elbaz) were based on a more compact array configuration, which resulted in a lower angular resolution data-set \citep{carlos2022a}. In this work, we use the low-resolution data-set, which has a sensitivity of 95.2~$\mu$Jy beam$^{-1}$ and an angular resolution of $1\farcs330\times~0\farcs935$ (synthesized beam full width at half maximum (FWHM) along
the major$\times$minor axis). This choice is motivated by our interest in detections and flux measurements, as opposed to resolving the extent of the sources.

\subsection{Sample selection}

In this research, we use the source catalog provided by CANDELS in GOODS-S (G13), that includes the redshifts, M$_\star$, SFR, and other SED-derived parameters for the galaxies. We select sources in the redshift range $1.0\leq z\leq3.0$, where G13 cataloged 18,459 galaxies (out of the full sample of 34,930 galaxies). While the lower limit is chosen given our interest in high redshift galaxies, especially at cosmic noon, where the gas mass density reaches its peak, the upper limit is chosen considering that the completeness at $3<z<4$ may be compromised. As pointed out in \citet{Merida2023} (M23 hereafter), $H$-band-based catalogs can be affected by significant selection effects when the $H$-band photometric point (i.e. the flux within the $F160W$ HST filter) lies bluewards of the Balmer break. This feature is shifted to 320~nm at $z=3$ (400~nm at $z=4$).

We focus on galaxies with M$_\star>10^8$~M$_\odot$, taking into account the G13 mass completeness limits. G13 compute this limit by looking for the most massive galaxies whose flux is equal to the faint limit of the sample, which evolves with redshift (\citealt{Fontana2004}, \citealt{Grazian2015}). The limits we report were obtained considering a 90\% completeness limit in the flux of $H=26$~mag, which corresponds to the limit computed for the shallow part of GOODS-S. Additionally, we only keep in our sample those sources with M$_\star<10^{11}$~M$_\odot$. The number density of the G13 sample sharply decreases towards $>$10$^{11}$~M$_\odot$, with only 14 objects within our redshift range. They all have M$_\star$ slightly above $\sim10^{11}$~M$_\odot$, right at the limit, so they cannot trace a higher mass bin than the one considered in this work (10$^{10-11}$~M$_\odot$, see Sec~\ref{sec:stacking} for more information on the divisions in redshift and M$_\star$ in further analysis). Additionally, considering the uncertainty in M$_\star$ for these galaxies, 8 out of the 14 sources could fall into the 10$^{10-11}$~M$_\odot$ range. If we include these 14 galaxies in the highest mass bin considered in this work, the f$_{\mathrm{gas}}$ barely change, showing only a variation of the order of a hundredth.

\begin{figure}
    \centering
    \includegraphics[width=8.5cm]{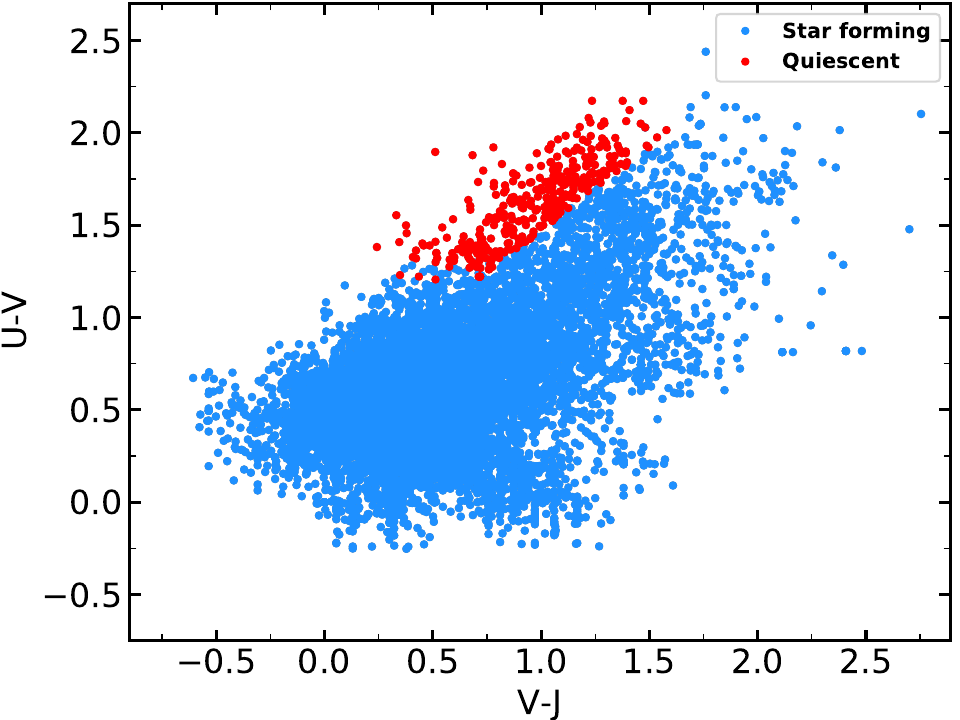}
    \caption{Rest-frame $U-V$ versus $V-J$ color for the G13 sample within $1\leq z\leq 3$ and $10^{8-11}$~M$_\odot$. In red we show the sources classified as quiescent according to this diagram. The star-forming sources are depicted in blue.}
    \label{fig:UVJ}
\end{figure}

We, moreover, restrict the sample to SFGs as indicated by the $UVJ$ diagram \citep{Whitaker2011}, which allows us to classify galaxies in quiescent or star-forming according to their rest-frame colors. This $UVJ$ selection guarantees that the M$_{\mathrm{gas}}$ we derive, based on stacking galaxies, is not biased to lower values because of the contribution of quiescent galaxies. In Fig.~\ref{fig:UVJ}, we show the $UVJ$ diagram depicting the galaxies that enter the selection and those that are discarded. All these criteria leave us with a sample of 15,236 sources.

Finally, we discard any source lying outside the GOODS-ALMA map coverage. This coverage is defined as the area where the noise is uniform across the map, excluding the edges of the outermost pointing, where there is no pointing overlap.
Our final sample is thus composed of 5,530 star-forming objects located at $1.0\leq z\leq3.0$, with stellar masses ranging 10$^{8-11}$~M$_\odot$. This sample shows typical uncertainties in the SED-derived parameters of 0.11 for the redshifts, 0.07~dex for the M$_\star$, and 0.05~dex for the SFRs.

Within this sample, we looked for the ALMA counterparts of our galaxies, as well as for galaxies in the vicinity of sources showing an ALMA counterpart, using the source catalog listed in \citet{carlos2022a}, GG22 hereafter. We select those galaxies closer than 5\arcsec\, to any object included in GG22. This 5\arcsec\, radius is chosen considering the growth curve of the low-resolution ALMA map point spread function (PSF) and the trade-off between the number of objects in the sample and the possible contamination of ALMA detected galaxies. This condition affects just $\sim$3\% of the objects in the sample. The effect of the exclusion of these individually detected sources and their neighbors in the photometry and further f$_{\mathrm{gas}}$ is discussed in Sec.~\ref{sec:stacking} and \ref{sec:fgas}. We will refer to the subsample obtained when excluding these galaxies as the undetected data-set hereafter.

Additionally, we also looked for counterparts of these sources in other ALMA-based catalogs, namely the ALMA twenty-six arcmin$^2$ survey of GOODS-S at one millimeter (ASAGAO; \citealt{Hatsukade2018}) and the ALMA Hubble Ultra Deep Field (ALMA-HUDF; \citealt{Dunlop2017}, \citealt{Hill2023}). In \citet{Yamaguchi2020}, they include a list of those ASAGAO sources that have an optical counterpart in the Fourstar Galaxy Evolution Survey (ZFOURGE; \citealt{Straatman2016}), and in \citet{Hill2023} they include the optical counterparts in G13 of the sources from the ALMA-HUDF catalog. Only 4 sources from the undetected data-set coincide with objects from ASAGAO and another 4 with sources from ALMA-HUDF.

We also investigated if any of these sources belonging to the undetected data-set shows significant emission at (sub-)millimeter wavelengths. We measured the photometry of each of these objects using the aperture photometry method provided by \texttt{photutils} from \texttt{Python}, selecting an aperture radius $r=0\farcs8$. Only 18 out of the undetected sources show a signal-to-noise ratio SNR$>$3. Only 3 out of these 18 galaxies show an SNR$>$3.5. When considering the SNR at the peak, only 1\% of the galaxies from the undetected data-set show an SNR$_{\mathrm{peak}}>3.5$ (including the latter 18 sources). Based on the analysis carried out in GG22, none of the sources is massive enough for the ALMA emission excess to be regarded as real and, therefore, they are indistinguishable from random noise fluctuations (see GG22 for more details). Given the low SNR of our sources, in order to look into the gas reservoirs of these galaxies we need to analyze stacked data.

\section{Properties of the sample and comparison with other catalogs} \label{sec:properties}

\begin{figure*}
    \includegraphics[width=16.5cm,height=14.5cm]{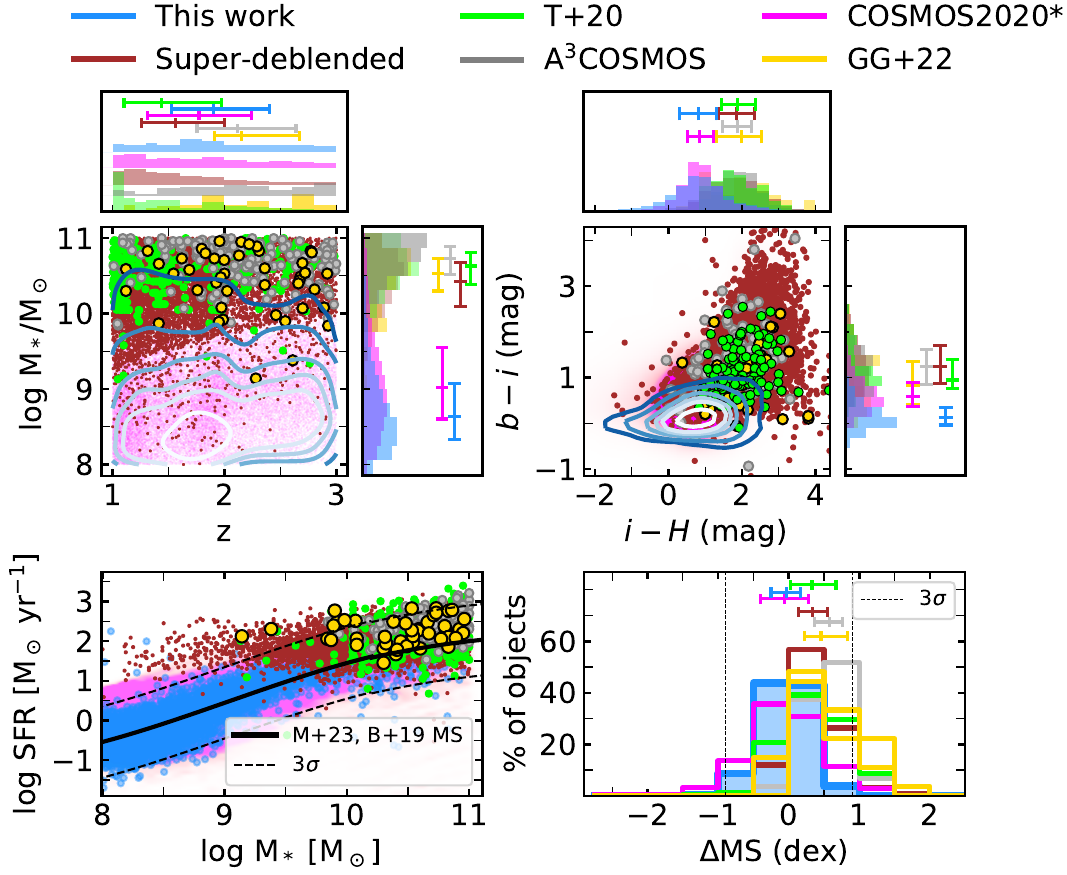}
    \centering{
    \includegraphics{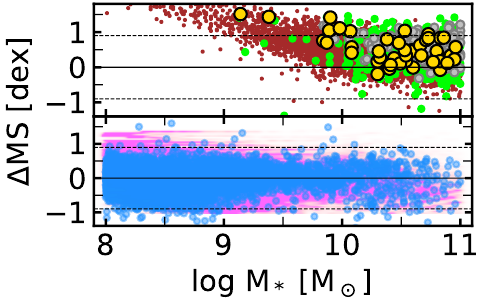}}
    \caption{From left to right and up and down we show: stellar mass vs. redshift, a color vs. color diagram based on $i-H$ and $b-i$, the star formation rate vs. stellar mass, histograms showing the distance to the main sequence in log scale, $\Delta$MS, and $\Delta$MS vs. stellar mass. We cut the comparison samples to only include galaxies within $1.0<z<3.0$ and having $10^{8-11}$~M$_\odot$.
    In blue, we represent our sample. The GG22 galaxies are identified in yellow. The \citet{Tacconi2020} and A$^3$COSMOS samples are shown in gray and green, respectively. The COSMOS2020$^*$ sample is represented in magenta and the galaxies from the super-deblended catalogs are displayed in maroon. In the stellar mass vs. redshift plot, the blue contours showing our sample enclose roughly 20\%, 50\%, 60\%, 70\%, 80\%, and 90\% of the data. In the color-color diagram, the contours roughly enclose the 10\%, 20\%, 40\%, 60\%, 80\%, and 90\% of the data. 
    In these two panels, histograms of the quantities there represented are also included, following the same color code. Quartiles are represented as horizontal segments in all the histograms. In the $z$-histograms shown in the first panel, we artificially elevate the baselines for the sake of clarity. The \citet{Merida2023} main sequence up to $10^{10}$~M$_\odot$ and the \citet{Barro2019} main sequence above $10^{10}$~M$_\odot$ are shown in the third panel as a solid black line. The dashed lines in the third, fourth, and fifth panels show the area enclosed within 3$\sigma$ with respect to the main sequence, based on the typical scatter reported in \citet{Speagle2014} ($\sim$0.30~dex). The last panel, showing $\Delta$MS vs. the stellar mass, is split for the sake of clarity, distinguishing between the super-deblended catalogs, \citet{Tacconi2020}, A$^3$COSMOS, and GG22 (top), and our sample and COSMOS2020$^*$ (bottom). The typical uncertainties for the redshifts, stellar mass, and SFRs of our galaxies are small, $\sim$0.11, 0.07~dex, and 0.05~dex, respectively. In the case of the $i-H$ and $b-i$ colors, these are 0.14~mag and 0.20~mag, respectively.}
    \label{fig:properties}
\end{figure*}

In this section, we compare the properties of the galaxies from our sample with those from other catalogs that were used by previous studies that also aimed at inferring the gas content of galaxies. In particular, we will refer to (i) the sources from the "super-deblended" catalogs, performed in GOODS-N and in the Cosmic Evolution Survey (COSMOS; \citealt{Scoville2007}), that were used to derive the \citet{Kokorev2021} scaling relation; (ii) the galaxies from the Automated ALMA Archive mining in the COSMOS field (A$^3$COSMOS), used to obtain the \citet{Liu2019b} scaling relation; (iii) the objects from \citet{Tacconi2020}, based on a compilation of individually detected galaxies from different surveys and also stacks, and used to derive their scaling relation; (iv) the galaxies from the COSMOS2020 catalog that lie in the A$^3$COSMOS footprint (COSMOS2020$^*$ hereafter), which is the sample the \citet{Wang2022} scaling relation is based on and, finally, (v) the sources from GG22. All these comparison samples are cut to only include galaxies within the same M$_\star$ and redshift intervals that we are considering in this work (see Sec.~\ref{sec:sample}). Later, in Sec.~\ref{sec:scaling}, we will refer to these same data-sets in the context of scaling relations.
In Fig.~\ref{fig:properties}, we show different diagrams that highlight the properties of the listed samples together with our data-set, based on G13. In Table~\ref{tab:comparison}, we summarize the information contained in Fig.~\ref{fig:properties}.

Given that some of these works do not report the values of the rest-frame colors of their sources, the  $i-H$ and $b-i$ colors allow us to build a diagram that works similarly to an $UVJ$, but using apparent magnitudes.
In Fig.~\ref{fig:properties}, for the panel showing this color vs. color diagram, we use the photometry measured within the $F435W$ ($b$), the $F775W$ ($i$), and the $F160W$ ($H$) bands from HST for our sample and for GG22. For A$^3$COSMOS and COSMOS2020$^*$, we use the photometry measured within the Subaru Prime Focus Camera (Suprime-Cam) $b$ band, the Hyper Suprime-Cam (HSC) $i$ band, and the UltraVISTA $H$ band. For the super-deblended catalogs and the \citet{Tacconi2020} data-set we also use the HST photometry, together with the Canada-France-Hawaii Telescope (CFHT) and Subaru observations in the absence of HST data. 

Fig.~\ref{fig:properties} also compares the position of the samples in the SFR vs M$_\star$ plane. In that panel, we include the M23 fit, defined for $1.5<z<2.0$, up to $10^{10}$~M$_\odot$, and the \citet{Barro2019} MS fit, B19 hereafter, above $10^{10}$~M$_\odot$. The distance of each point to the MS is re-scaled to its corresponding redshift. In the fourth panel, we show the difference with respect to the MS for each galaxy in these samples, $\Delta$MS ($\Delta$MS = log SFR$-$log SFR$_\mathrm{MS}$, with $\Delta$MS = 0 being equivalent to $\delta$MS = SFR/SFR$_\mathrm{MS}$ = 1). We use M23 to calculate the $\Delta$MS of galaxies with M$_\star$/M$_\odot<10^{10}$ and B19 for sources with higher stellar masses, given that M23 focuses on the low-mass end of the MS.

It is important to mention that for our G13-based sample, \citet{Tacconi2020}, and COSMOS2020$^*$, the SFR were computed following the ladder technique \citep{Wuyts2011}, which combines SFR indicators at UV, mid-infrared (MIR) and FIR. 
For the galaxies from the super-deblended catalogs, the SFR was computed from the
integrated IR luminosity (LIR) using the \citet{Daddi2010} relation.
The SFRs for A$^3$COSMOS were computed from the IR luminosity using the \citet{Kennicutt1998} calibration. For the GG22 galaxies, the SFRs were calculated as the sum of the SFR$_{\mathrm{IR}}$ (using the \citealt{Kennicutt1998} calibration) and the SFR$_{\mathrm{UV}}$ (using the \citealt{Daddi2004} calibration). We checked that only 219 of our galaxies (4\%, with only 10 galaxies with M$_\star>10^{10}$~M$_\odot$) are detected in the MIR and/or FIR using \textit{Spitzer} MIPS and \textit{Herschel} PACS and SPIRE. This means that the SFRs of our sample come mostly from the UV emission.

\subsection{This work: the G13-based data-set}

The sample evenly populates the redshift range considered in this work (first panel of the first row in Fig.~\ref{fig:properties}), with median and quartiles $z=1.9_{1.5}^{2.4}$. The low-mass coverage of G13 allows us to reach down to $10^{8-9}$~M$_\odot$ (log (M$_\star$/M$_\odot$)=$8.6_{8.3}^{9.1}$). In terms of the optical colors (second panel of the first row), our sample shows typical values of 0.81$_{0.29}^{1.31}$~mag for $i-H$ and 0.12$_{-0.04}^{0.35}$~mag for $b-i$.

The position of our galaxies in the SFR vs M$_\star$ plane (first panel of the second row) is compatible with the MS, with only a minor population of galaxies above or below three times the typical scatter ($\sim0.04$\% and $\sim1$\% of the galaxies above/below 3$\sigma$, respectively, with $\sigma$ being $\sim\,$0.3~dex according to \citealt{Speagle2014}). The median $\Delta$MS (second panel of the second row) of our galaxies is $\Delta$MS~$=-0.03_{-0.25}^{0.17}$~dex. If we check the position of our galaxies with respect to the MS according to M$_\star$ (third row), we see that this typical $\Delta$MS is maintained over the whole M$_\star$ range, including the high-mass end, where most of our comparison samples are located, thus making it difficult to see our galaxies in the SFR vs M$_\star$ plane above $10^{9.5}$~M$_\odot$.

\subsection{Comparison data-sets}

\begin{itemize}
    \item "Super-deblended" catalogs
\end{itemize}

The "super-deblended" catalogs (\citealt{Liu2018}, \citealt{Jin2018}), performed in GOODS-N and COSMOS and constructed using FIR and sub-millimeter images, use the prior positions of sources from deep
\textit{Spitzer}/IRAC and Very Large Array (VLA) 20~cm observations to obtain the photometry of blended FIR/sub-millimeter sources. They also employ the SED information from shorter wavelength photometry as a prior to subtract lower redshift objects. In the case of the COSMOS super-deblended catalog, the authors additionally select a highly complete sample of priors in the $Ks$-band from the UltraVista catalogs. Apart from selecting those galaxies satisfying our redshift and M$_\star$ cuts, we only keep those galaxies showing an SNR~$>3$ in at least 3 FIR to sub-millimeter bands from 100~$\mu$m to 1.2~mm, following \citet{Kokorev2021}. The optical photometry of these galaxies is obtained by looking for possible optical counterparts in the CANDELS catalog performed in GOODS-N (B19) and in the COSMOS2020 catalog \citep{Weaver2022}.

The median redshift and quartiles of the galaxies satisfying our selection criteria are $z=1.6_{1.1}^{2.0}$, with a higher concentration of lower redshift galaxies compared to our sample, and in line with T20. This data-set is biased towards more massive galaxies than our sample (log M$_\star$/M$_\odot$ = 10.4$_{10.1}^{10.7}$), $\sim1.8$~dex more massive than our galaxies. Their optical $i-H$ and $b-i$ colors are redder than the ones traced by our sources ($i-H=1.85_{1.35}^{2.34}$~mag and $b-i=1.24_{0.88}^{1.71}$~mag, respectively), $\sim1$~mag redder in both colors. 

In terms of the position of these galaxies with respect to the MS, these sources show $\Delta$MS values compatible with being MS galaxies, showing $\Delta$MS = $0.34_{0.14}^{0.55}$~dex. This value corresponds though to a more star-forming data-set, more compatible with the upper envelope of the MS, given the typical scatter. 6\% of the galaxies show values $>$3$\sigma$. If we examine the evolution of $\Delta$MS with M$_\star$, we see that below $10^{10}$~M$_\odot$, the sample gets increasingly star-forming, with most of the sample below 10$^9$~M$_\odot$ surpassing $\Delta$MS=1. 
 
\begin{itemize}
  \item A$^3$COSMOS
\end{itemize}

\begin{table}
\setlength{\tabcolsep}{2.05pt} 
\renewcommand{\arraystretch}{1.5}
\caption{Summary of the properties of our data-set and the comparison samples shown in Fig.~\ref{fig:properties}, and described along Sec.~\ref{sec:properties}. The comparison samples are limited to the redshifts and stellar masses studied in this work. The values here shown correspond to the median and first and third quartiles. The names between brackets refer to the scaling relations that are based on each data-set (see Sec.~\ref{sec:scaling}). W22 refers to the \citet{Wang2022} relation, K21 to the relation from \citet{Kokorev2021}, and L19 to that of \citet{Liu2019b}. The T20 data-set was used in that same work to obtain their scaling relation.}             
\label{tab:comparison}      
\centering          
\small 
\begin{tabular}{c c c c c c}        
\hline\hline                 
 & z & log M$_\star$ & $i-H$ & $b-i$ & $\Delta$MS\\ & & (M$_\odot$) & (mag) &(mag) &(dex) \\     
\hline                        
\multicolumn{6}{c}{\textbf{Mass-complete samples}}\\
\hline
\textbf{This work} &1.9$_{1.5}^{2.4}$ &8.6$_{8.3}^{9.1}$&0.81$_{0.29}^{1.31}$ &0.12$_{-0.04}^{0.35}$ & $-0.03_{-0.25}^{0.17}$  \\
\hline
\textbf{COSMOS2020$^*$ (W22)}&1.8$_{1.3}^{2.2}$ &9.0$_{8.6}^{9.6}$&0.85$_{0.51}^{1.21}$ &0.59$_{0.37}^{0.89}$ & $-0.06_{-0.39}^{0.28}$  \\
      \hline
\multicolumn{6}{c}{\textbf{Individual (sub-)mm detections + stacks}}\\
\hline
\textbf{T20}&1.4$_{1.1}^{2.0}$ &10.6$_{10.4}^{10.8}$&1.87$_{1.45}^{2.36}$ &0.95$_{0.76}^{1.40}$ & $0.33_{0.02}^{0.67}$  \\
      \hline

\multicolumn{6}{c}{\textbf{Individual (sub-)mm detections}}\\
\hline
\textbf{Super-deblended (K21)}&1.6$_{1.1}^{2.0}$ &10.4$_{10.1}^{10.7}$&1.85$_{1.35}^{2.34}$ &1.24$_{0.88}^{1.71}$ & $0.34_{0.14}^{0.55}$  \\
      \hline
     \textbf{A$^3$COSMOS (L19)}&2.1$_{1.8}^{2.6}$ &10.7$_{10.5}^{10.9}$&1.89$_{1.46}^{2.27}$ &1.24$_{0.86}^{1.61}$ & $0.58_{0.36}^{0.76}$  \\
      \hline
     \textbf{GG22}&2.2$_{1.9}^{2.7}$ &10.5$_{10.3}^{10.7}$&1.98$_{1.30}^{2.54}$ &0.82$_{0.41}^{1.36}$ & $0.46_{0.22}^{0.83}$  \\
\hline                                   
\end{tabular}
\end{table}

The A$^3$COSMOS dataset \citep{Liu2019a} contains $\sim$700 galaxies ($0.3 < z < 6$) with high-confidence ALMA detections in the (sub-)millimeter continuum. It consists of a blind extraction, imposing an SNR$_\mathrm{peak}>5.40$, and on a prior-based extraction, using the known positions of sources in the COSMOS field, cutting the final sample to SNR$_\mathrm{peak}>4.35$. We extract the photometry of these sources from the COSMOS2020 catalog.

The A$^3$COSMOS galaxies with redshifts and M$_\star$ in common with this work are mainly located at higher redshifts ($z=2.11_{1.75}^{2.64}$) compared to our galaxies, and are also biased towards more massive objects (log (M$_\star$/M$_\odot$) = $10.7_{10.5}^{10.9}$), $\sim2$~dex more massive than our sources in this case. They also display redder optical colors, with values of 1.89$_{1.46}^{2.27}$~mag for $i-H$ and 1.24$_{0.86}^{1.61}$~mag for $b-i$, $\sim$1~mag redder in both colors than our sample. 

According to their position in the SFR vs. M$_\star$ plane, these objects are also compatible with the MS, but, as well as the galaxies from the super-deblended catalogs, T20, and GG22, they are located in the upper envelope, showing values nearly 2 times the typical scatter ($\Delta$MS = 0.58$_{0.36}^{0.76}$~dex, with 13\% of the galaxies above 3$\sigma$).

\begin{itemize}
    \item \citet{Tacconi2020}
\end{itemize}

The \citet{Tacconi2020} sample (T20 hereafter) is based on the existing literature and ALMA archive detections for individual galaxies and stacks. It consists of 2,052 SFGs. 858 of the measurements are based on CO detections, 724 on FIR dust measurements, and 470 on $\sim$1~mm dust measurements. We extract their photometry looking for the counterparts of the individual objects in the CANDELS catalogs performed in the different cosmological fields, using the catalogs already specified together with the \citealt{Stefanon2017} catalog for the Extended Groth Strip (EGS; \citealt{Davis2007}), and \citealt{Galametz2013} for the Ultra Deep Survey (UDS; \citealt{Lawrence2007}, \citealt{Cirasuolo2007}). It is however true that, since part of their sample is based on stacking, our results regarding the colors will only reflect the nature of the individual detections that make up the sample.

We see that the \citet{Tacconi2020} galaxies meeting our redshift and M$_\star$ criteria are centered at $z=1.4_{1.1}^{2.0}$, in line with the super-deblended sample. In terms of M$_\star$, this data-set is made up mostly of massive objects (log (M$_\star$/M$_\odot$) = $10.6_{10.4}^{10.8}$), 2~dex more massive than our sample. According to the optical colors, this sample traces redder values of $i-H$ and $b-i$, typically $1.87_{1.45}^{2.36}$~mag and $0.95_{0.76}^{1.40}$~mag for each of these colors. This is more than 1~mag redder in $i-H$ and $\sim0.8$~mag redder in $b-i$. 

These galaxies are more star-forming than our sources, showing $\Delta$MS = 0.33$_{0.02}^{0.67}$~dex, which is compatible with them being in the upper envelope of the MS (13\% of the galaxies are located above 3$\sigma$).

\begin{itemize}
    \item COSMOS2020$^*$
\end{itemize}

The COSMOS2020 catalog comprises 1.7 million sources across the 2 deg$^2$ of the COSMOS field, $\sim$966,000 of them measured with all available broad-band data. Compared to COSMOS2015 \citep{Laigle2016}, it reaches the same photometric redshift precision at almost one magnitude deeper. It goes down to $10^{8.43}$~M$_\odot$ at $z=1$ with 70\% completeness ($10^{9.03}$~M$_\odot$ at $z=3$). We keep those galaxies that lie within the A$^3$COSMOS footprint, which we will call COSMOS2020$^*$, consisting of 207,129 objects. This sample is not biased towards ALMA-detected galaxies, CO emitters, or high-mass systems, which makes it more similar to our data sample.

The median redshift of the galaxies within our redshift and M$_\star$ intervals is $z=1.77_{1.31}^{2.24}$, comparable to the values we retrieve for our sample. The COSMOS2020$^*$ data-set shows a typical M$_\star$ of log (M$_\star$/M$_\odot$) = $9.0_{8.6}^{9.6}$, $\sim$0.5~dex more massive than our sample. According to the optical colors, these sources show similar $i-H$ colors ($0.85_{0.51}^{1.21}$~mag) to our galaxies, and around $\sim0.50$~mag redder colors of $b-i$ ($0.59_{0.37}^{0.89}$~mag). 

These objects are located well within the MS typical scatter, with a $\Delta$MS very similar to the one we obtain for our data-set ($\Delta$MS = $-0.06_{-0.39}^{0.28}$~dex, with 4\% of the galaxies above 3$\sigma$ and 7\% of the galaxies below 3$\sigma$).

\begin{itemize}
    \item GG22
\end{itemize}

GG22 presented an ALMA blind survey at 1.1 mm and built a bona fide sample of 88 sources, comprising mostly massive dusty star-forming galaxies. Half of them are detected with a purity of 100\% with a SNR$_\mathrm{peak}>5$ and half of them with 3.5~$\le$~SNR$_\mathrm{peak}\le$~5, aided by the \textit{Spitzer}/IRAC and the VLA prior positions. We retrieve the optical fluxes of the GG22 ALMA-selected galaxies from ZFOURGE.

The GG22 sources compatible with our redshifts and M$_\star$ cuts are also biased towards high redshifts compared to our sample, similarly to A$^3$COSMOS ($z=2.15_{1.91}^{2.67}$). As well as the super-deblended data-set, T20, and A$^3$COSMOS, GG22 is mainly made up of massive galaxies, $\sim$2~dex more massive than our objects (log (M$_\star$/M$_\odot$) = $10.5_{10.3}^{10.7}$). Their optical colors are also redder than the ones showed by our sample, with median and quartiles being $1.98_{1.30}^{2.54}$~mag for $i-H$ ($\sim$1~mag redder) and $0.82_{0.41}^{1.36}$~mag for $b-i$ (0.7~mag redder). 

These galaxies are also MS objects but, as well as the sources from the super-deblended data-set, T20, and A$^3$COSMOS, they are more star-forming than our sources ($\Delta$MS = 0.46$_{0.22}^{0.83}$~dex), located above the typical scatter of the MS (20\% of the galaxies above 3$\sigma$).

\subsection{Comparison remarks}

The main differences between our data-set and the comparison samples, with the exception of COSMOS2020$^*$, are the blue optical colors of our galaxies, their low M$_\star$ coverage, and their closer proximity to the MS. However, the latter results concerning the blue optical colors of our galaxies can be a consequence of mixing different redshifts and M$_\star$ when producing the color-color diagram. We thus decided to cut it in 0.5 redshift bins and select galaxies with $>10^{10}$~M$_\odot$, which allows a direct comparison with the other catalogs. Below $10^{10}$~M$_\odot$, as highlighted in different panels in Fig.~\ref{fig:properties}, we lack sources to compare. The red colors of the comparison samples are thus due to $>10^{10}$~M$_\odot$ systems. 

\begin{figure*}
    \centering \includegraphics[width=15.8cm,height=17.3cm]{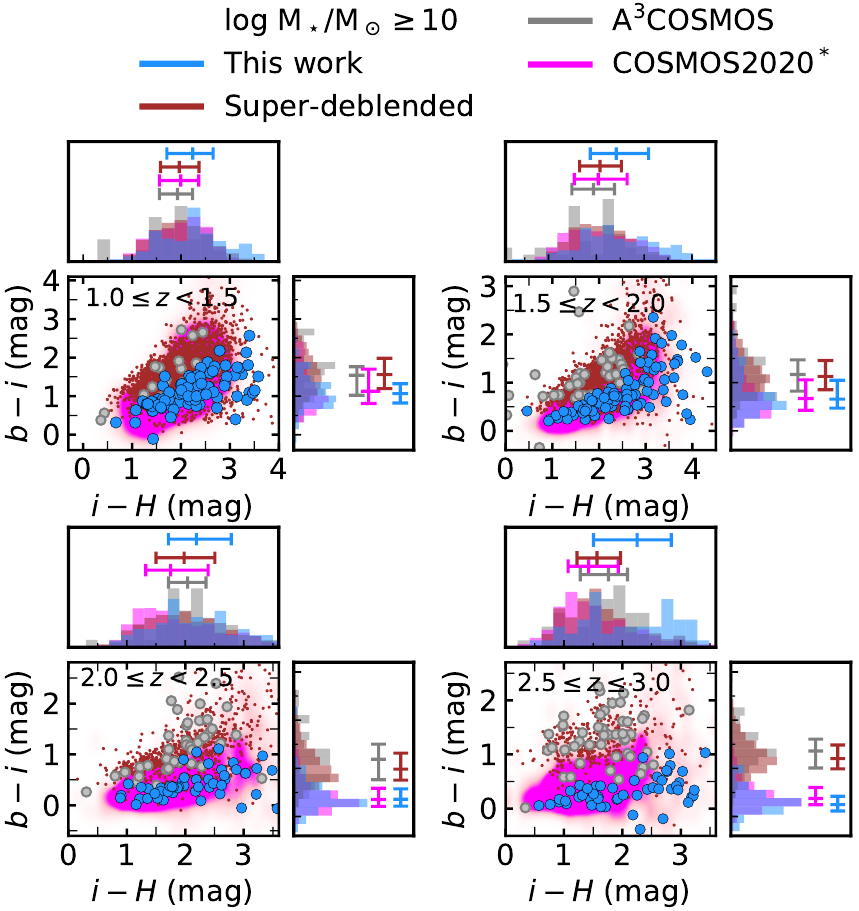}
    \caption{Color vs. color diagram based on $i-H$ and $b-i$ in different redshift bins. We only include galaxies with M$_\star\geq10^{10}$~M$_\odot$ within our data-set, the super-deblended catalogs, A$^3$COSMOS, and COSMOS2020$^*$. See Fig.~\ref{fig:properties} for the description of the color codes and markers here shown.}
    \label{fig:properties_2}
\end{figure*}

In Fig.~\ref{fig:properties_2}, we show the color-color diagram included in Fig.~\ref{fig:properties} divided into different redshift bins. We only show the super-deblended, A$^3$COSMOS and COSMOS2020$^*$ galaxies as comparison data-sets since the number of objects in each redshift bin included in these catalogs still provides the means to obtain meaningful number statistics to compare with.

When restricting our sample to galaxies with $>10^{10}$~M$_\odot$, the difference in $i-H$ diminishes and we retrieve similar values to those obtained for the comparison samples. We get $2.23_{1.71}^{2.66}$~mag at $1.0\leq z <1.5$, and $2.25_{1.51}^{2.84}$~mag at $2.5\leq z\leq3.0$ for our data-set.

For the $b-i$ color, we trace bluer values than the super-deblended catalogs and A$^3$COSMOS while getting similar results to COSMOS2020$^*$. The difference between the set of our sample and COSMOS2020$^*$, and the super-deblended catalogs and A$^3$COSMOS increases with redshift, and the color gets bluer as well. We obtain
$1.07_{0.82}^{1.32}$~mag at $1.0\leq z <1.5$ ($0.25_{0.12}^{0.42}$~mag at $2.5\leq z\leq3.0$) according to our data-set, compared to $1.56_{1.19}^{1.98}$~mag at $1.0\leq z <1.5$ ($1.20_{0.99}^{1.48}$~mag at $2.5\leq z\leq3.0$) for the super-deblended catalogs, and $1.54_{1.02}^{1.76}$~mag at $1.0\leq z <1.5$ ($1.36_{1.01}^{1.61}$~mag at $2.5\leq z\leq3.0$) for A$^3$COSMOS. The similarity with COSMOS2020$^*$ and the discrepancy with the super-deblended catalogs and A$^3$COSMOS in this color are expected. The COSMOS2020$^*$ includes all the galaxies at these M$_\star$, regardless of their flux at (sub-)millimeter wavelengths, hence being mass-complete at $10^{10}$~M$_\odot$, similarly to our sample. On the contrary, the super-deblended catalogs use prior positions from deep \textit{Spitzer}/IRAC and VLA observations, and the A$^3$COSMOS only considers sources with high-confidence ALMA detections, which is translated to redder colors of $b-i$ and higher dust obscurations. Our galaxies show median optical extinctions, A(V), ranging from 1.03-1.71~mag, smaller as we increase in redshift, whereas these numbers are 2.08-2.28~mag for A$^3$COSMOS.

\section{Stacking analysis and flux measurements} \label{sec:stacking}

In order to study the gas content of our galaxies, we stack the emission of objects similar to each other. We group galaxies according to (1) redshift and (2) log M$_\star$. We distinguish:

\begin{itemize}
  \item 4 redshift bins: $1.0\leq z<1.5$, $1.5\leq z<2.0$, $2.0\leq z<2.5$, and $2.5\leq z\leq3.0$
  \item 3 M$_\star$ bins, 8$\leq$log M$_\star/$M$_\odot<$9, 9$\leq$log M$_\star/$M$_\odot<$10, 10$\leq$log M$_\star/$M$_\odot \leq$11
\end{itemize}

These divisions in redshift and stellar mass are chosen as a result of an estimation used to evaluate and maximize the probability of obtaining detections according to different combinations of redshift and M$_\star$ intervals. The estimation is based on the depth of the observations and the previous knowledge about the gas reservoirs in galaxies as given by the scaling relations derived in other works (see Sec.~\ref{sec:scaling}). If we consider the expected gas fractions provided by these relations and use the $\delta_{\mathrm{GDR}}$ approach (see Sec.~\ref{sec:intro} and \ref{sec:fgas}), we can calculate the typical flux density that corresponds to those gas fractions and we can roughly infer the number of objects necessary to obtain a measurement with SNR~$>3$. For this, we quantify the relation $\sigma$ $\;\propto 1/\sqrt{N}$ (with $\sigma$ being the resulting noise in the stacked map and $N$ the number of objects) considering different combinations of redshift and M$_\star$ bins and obtain that, for getting a measurement (SNR~$>3$) at 10~$\leq$~log M$_\star/$M$_\odot \leq$~11, just a few objects ($<$~10) are required. For the 9~$\leq$~log M$_\star/$M$_\odot<$~10 bin, we require a number of objects of the order of hundreds. Finally, for the 8~$\leq$~log M$_\star/$M$_\odot<$~9 bin, we would need tens of thousands of objects to reach the necessary depth according to our current knowledge of the gas reservoirs in galaxies. We check that the adopted redshift division guarantees these numbers for the 9~$\leq$~log M$_\star/$M$_\odot<$~10 and 10~$\leq$~log M$_\star/$M$_\odot \leq$~11 mass bins, while for the 8~$\leq$~log M$_\star/$M$_\odot<$~9 bin, we lack objects, regardless of how we divide in redshift, what already warns that the probability to obtain a measurement in this M$_\star$ bin is very low. This estimation does not, however, ensure that we are obtaining measurements for the two remaining bins, given that scaling relations are not calibrated for the kind of objects we are considering in this work, but still can be used as a starting point.

After defining the bins, we stack 50$\times$50 arcsec$^2$ (1,000$\times$1,000~pixel$^2$) cutouts within the low-resolution ALMA mosaic, centered at each source and using the coordinates of the centroids provided by G13. Before the stacking, we corrected these centroids for a known offset between the HST and ALMA data, reported in different studies (e.g. \citealt{Dunlop2017}, \citealt{Franco2018}). We apply the correction from \citet{Whitaker2019}, that corresponds to $\delta$R.A.(deg) = (0.011$\pm$0.08)/3600 and $\delta$decl.(deg) = ($-0.26\pm0.10)/3600$. We opted for median stacking galaxies instead of mean stacking them. This choice is motivated by our aim to provide an estimate of the gas reservoirs of the bulk of the SFG population, not biased towards bright sources. Additionally, this method allows us to get closer to the detection threshold in the case of the $10^{9-10}$~M$_\odot$ bin; the use of mean stacking reports lower SNR in 3 of the 4 redshift bins.

Despite our choice, we computed the fluxes and further physical parameters from both, mean stacking and median stacking measurements. In Sec.~\ref{sec:fgas} we discuss quantitatively the effects introduced by mean or median stacking the galaxies in the gas content, and in Appendix~\ref{app:mean_stack} we include analogs of some of the figures and tables appearing in this paper showing the results obtained using mean stacking.

We checked that the centroids computed using the stacked emission in ALMA are compatible with those provided by G13, based on the HST imaging, within $0\farcs06$.
Following GG22, the photometry is calculated within an aperture of $r=0\farcs8$. This radius provides the optimal trade-off between total flux retrieval and total SNR for the GG22 sample.
We then apply the corresponding aperture correction by dividing this flux density by that enclosed within the synthesized dirty beam (normalized to its maximum value) using the same aperture radius (see \citealt{carlos2022a} for more details). This aperture correction is $\sim$1.67 for $r=0\farcs8$.

When the SNR~$<3$, we calculate an upper limit for the flux density based on the surrounding sky emission in the stacked image by placing 10,000 $r=0\farcs8$ apertures at random positions across a 20$\times$20~arcsec$^2$ cutout centered at the stacked source. We measure the photometry within each of these apertures and produce a histogram with all the values, fitting the resulting Gaussian distribution leftwards to the peak, to skip the possible emission of the source. We compute the upper limit as 3 times the standard deviation of the fit. We also checked that this approach is compatible with the standard deviation obtained by iteratively drawing $N$ (equal to the number of sources in the stack) empty positions along the mosaic, stacking them, and measuring the flux within a $r=0\farcs8$ aperture. The compatibility of the two methods is the result of the noise uniformity along the map.

If SNR~$>3$ within the aperture we repeat the measurement using an aperture radius $r=1\arcsec$. We checked that this larger radius allows us to optimize the flux/SNR gain/loss, recovering $\sim$7$\%$ more flux. The aperture correction for $r=1\arcsec$ is $\sim$1.28. The uncertainty associated with the measurements is the result of the combination of the error of the stacked data (which is used to compute the SNR) and the uncertainty linked to the underlying distribution of sources that contribute to the stack. The former component is calculated by placing 10,000 $r=1\arcsec$ apertures at random positions across the 50$\times$50~arcsec$^2$ stacked cutout. We measure the photometry within each aperture and fit the histogram leftwards to the peak, as done in the calculation of the upper limits in the previous case. The standard deviation provided by this fit is taken as the uncertainty. The uncertainty associated with the underlying distribution is computed via bootstrapping, considering the standard deviation of the bootstrap samples.

\begin{figure*}
    \centering
   \includegraphics[width=16.1cm, height=23.5cm]{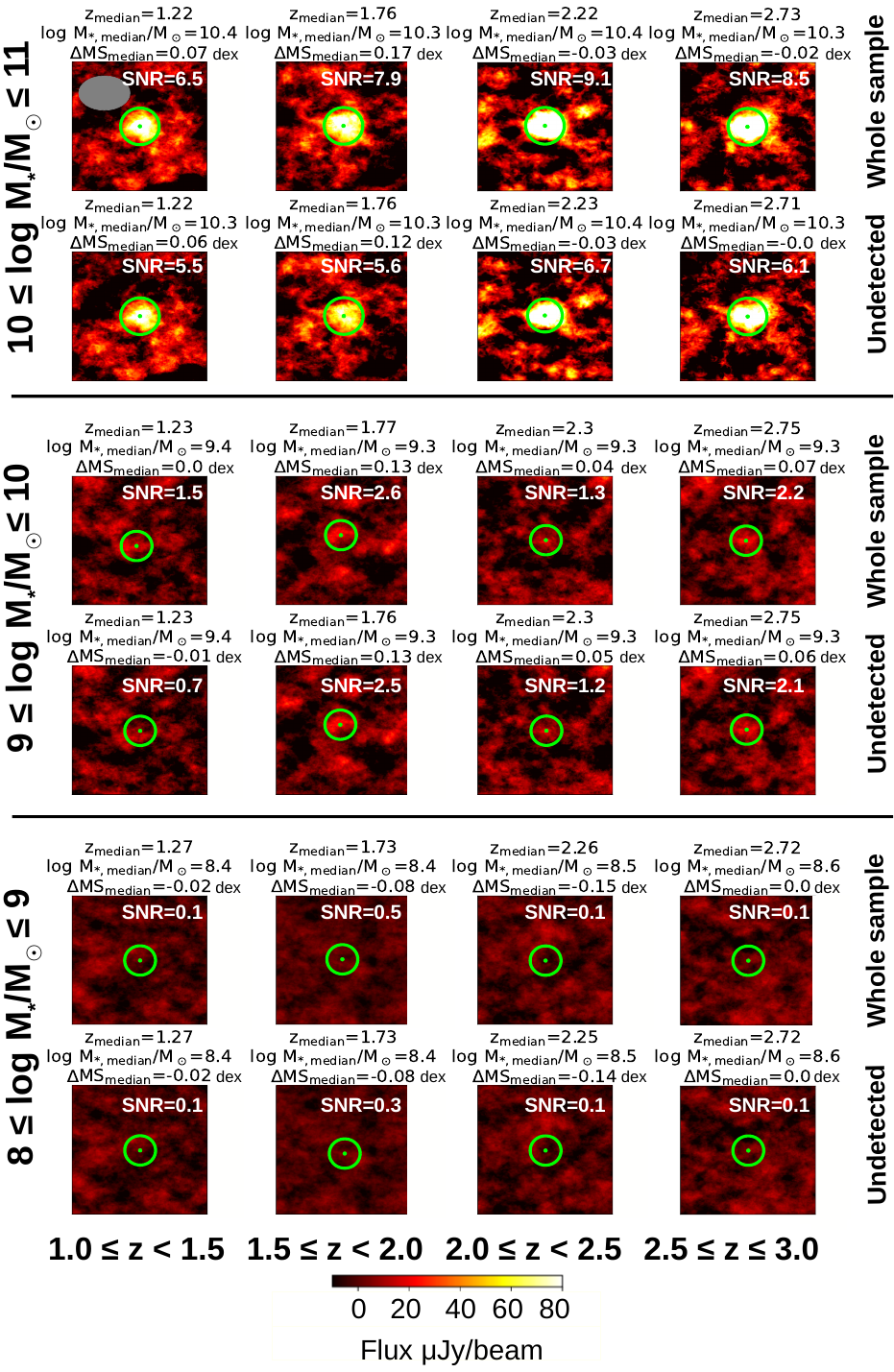}
    \caption{Cutouts of 7$\times$7 arcsec$^2$ of the ALMA low-resolution map showing the median stacked galaxies in each redshift and mass bin. For each bin, we include cutouts that correspond to the whole sample and the undetected data-set, as defined in Sec.~\ref{sec:sample}. The apertures used to measure the photometry are displayed in green. The size of the beam is shown in the first panel of the figure in gray. The flux densities and the corresponding uncertainties for each stacked galaxy are included in Table~\ref{tab:properties1}.}
    \label{fig:mass_bins}
\end{figure*}

In Fig.~\ref{fig:mass_bins} we show an image of all the stacks. In Table~\ref{tab:properties1} we list the flux densities we measure, together with the derived uncertainties. In both, Fig.~\ref{fig:mass_bins} and Table~\ref{tab:properties1}, we also include the results for the undetected data-set, defined in Sec.~\ref{sec:sample}. In Sec.~\ref{sec:fgas}, we discuss the effects of the inclusion or exclusion of the GG22 sources and their neighbors in the f$_{\mathrm{gas}}$.
For both, the whole sample and the undetected data-set, we obtain SNR~$>3$ flux density measurements for $10^{10-11}$~M$_\odot$ (high-mass bin) at all redshifts. The flux density enclosed within this M$_\star$ bin increases towards higher redshifts. For the intermediate-mass ($10^{9-10}$~M$_\odot$) and the low-mass bins ($10^{8-9}$~M$_\odot$), we provide 3$\sigma$ upper limits. In the case of the intermediate-mass bin, we obtain a signal close to our SNR threshold at $1.5\leq z<2.0$, with an SNR = 2.6 (SNR = 2.5 for the undetected data-set), and again at $2.5\leq z\leq3.0$, with an SNR = 2.2 (SNR = 2.1 for the undetected data-set). Given the lack of detection in the lower mass bins, we tested whether regrouping the galaxies, stacking all sources within 10$^{8-10}$~M$_\odot$, would allow the threshold to be exceeded. We did this for each redshift bin, and also including all galaxies at $1\leq z\leq2$ and $2<z\leq 3$. However, these tests do not report any detections; the galaxies in the low-mass bin dominate the emission of the stack.

\begin{table*}
\setlength{\tabcolsep}{12pt} 
\renewcommand{\arraystretch}{1}
\centering
\caption{Summary of the photometry derived in this work. We show the flux densities for our sample, and a subsample excluding the \citet{carlos2022a} sources and those from \citet{Guo2013} which are closer than 5\arcsec\, to them, as explained in Sec.~\ref{sec:sample}. We split the total uncertainty into its two components: the error due to the stacked data and the uncertainty linked to the underlying distribution of the individual galaxies. The total uncertainty is obtained by summing the previous error contributions in cuadrature. The absence of uncertainty denotes an upper limit at the 3$\sigma$ level.}
\label{tab:properties1}
\begin{tabular}{lllllll}
\hline\hline
\raggedleft
z bin&log M$_\star$ bin&N$_{\mathrm{obj}}$&Flux density &erFlux$_{\mathrm{stack}}$&erFlux$_{\mathrm{ind}}$&erFlux$_{\mathrm{total}}$\\&(M$_\odot$)& &($\mu$Jy)&($\mu$Jy)&($\mu$Jy)&($\mu$Jy)\\
\hline
& \textbf{Whole}& \textbf{sample} &&\\
\hline
        &$8\leq$~log M$_\star<$~9 &910 &$<$~17.31 & -&- &- \\
      1.0~$\leq z<$~1.5&$9\leq$~log M$_\star<$~10 & 299& $<$~25.88& - &- &- \\
     &$10\leq$~log M$_\star\leq$~11 & 92 & 132.10&20.36 &21.55 &29.65\\
\hline
       &$8\leq$~log M$_\star<$~9 &1423 &$<$~12.70 & -  &-&-\\
      1.5~$\leq z<$~2.0&$9\leq$log M$_\star<$10 & 353& $<$~26.13& -  &- &-\\
     &$10\leq$~log M$_\star\leq$~11 & 96 & 148.31&18.67 &24.49&30.79\\
\hline
     &$8\leq$~log M$_\star<$~9 &881 &$<$~14.99 & -  &-&-\\
    2.0~$<z<$~2.5&$9\leq$~log M$_\star<$~10 & 274& $<$~25.94& - &-&- \\
     &$10\leq$~log M$_\star\leq$~11 & 54 & 219.30&24.01 &25.33&34.90 \\
\hline
     &$8\leq$~log M$_\star<$~9 &751 &$<$~15.71 & -&- &-\\
     2.5~$\leq z\leq$~3.0&$9\leq$log M$_\star<$10 & 340& $<$~26.38& - &-&-  \\
     &$10\leq$~log M$_\star\leq$~11 & 57 & 211.26&24.79 &47.62 &53.69\\
\hline
& \textbf{Undetected}& \textbf{sources} &&\\
\hline
        &$8\leq$~log M$_\star<$~9 &890 &$<$~16.83 & - &- &-\\
      1.0~$\leq z<$~1.5&$9\leq$~log M$_\star<$~10 & 285& $<$~25.71& -  &-&- \\
     &$10\leq$~log M$_\star\leq$~11 & 85 & 115.97&20.91 &20.84&29.52 \\
\hline
       &$8\leq$~log M$_\star<$~9 &1387 &$<$~11.75 & - &- \\
      1.5~$\leq z<$~2.0&$9\leq$~log M$_\star<$~10 & 338& $<$~25.57& - &- &- \\
     &$10\leq$~log M$_\star\leq$~11 & 84 & 110.66&19.72&22.06 &29.59 \\
\hline
     &$8\leq$~log M$_\star<$~9 &854 &$<$~14.64 & -  &-&-\\
    2.0~$<z<$~2.5&$9\leq$~log M$_\star<$~10 & 265& $<$~30.55& -  &-&-\\
     &$10\leq$~log M$_\star\leq$~11 & 46 & 165.87&24.83 &29.67 &38.69\\
\hline
     &$8\leq$~log M$_\star<$~9 &732 &$<$~14.86 & - &-&-\\
     2.5~$\leq z\leq$~3.0&$9\leq$~log M$_\star<$~10 & 325& $<$~27.91& - &- &- \\
     &$10\leq$~log M$_\star\leq$~11 & 47 & 165.93&27.05  &31.84&41.78\\
\hline
\end{tabular}
\end{table*}

The use of a certain aperture radius in our measurements, in this case, $r=1.0$~\arcsec, involves some flux loss. Departure from a point-like source may involve an additional flux correction based on the galaxy morphology (see \citealt{Blanquez-Sese2023}). We consider two size estimations: the size of the dust component as prescribed by GG22 and the size of the stellar component as measured and reported in G13, based on $H$-band data.

As pointed out in several studies, the dust component is usually more concentrated than the stellar one (e.g. \citealt{Kaasinen2020}, \citealt{Tadaki2020}, \citealt{carlos2022a}, \citealt{Liu2023}). However, it is currently uncertain if our stacks, based on a mass-complete sample including faint objects, follow the latter statement, given that previous size estimations of the dust component rely on individual detections of bright objects at (sub-)millimeter wavelengths. Due to this, we also include a size estimation based on the stellar component.

\begin{figure}
    \centering
\includegraphics[width=8cm,height=6.1cm]{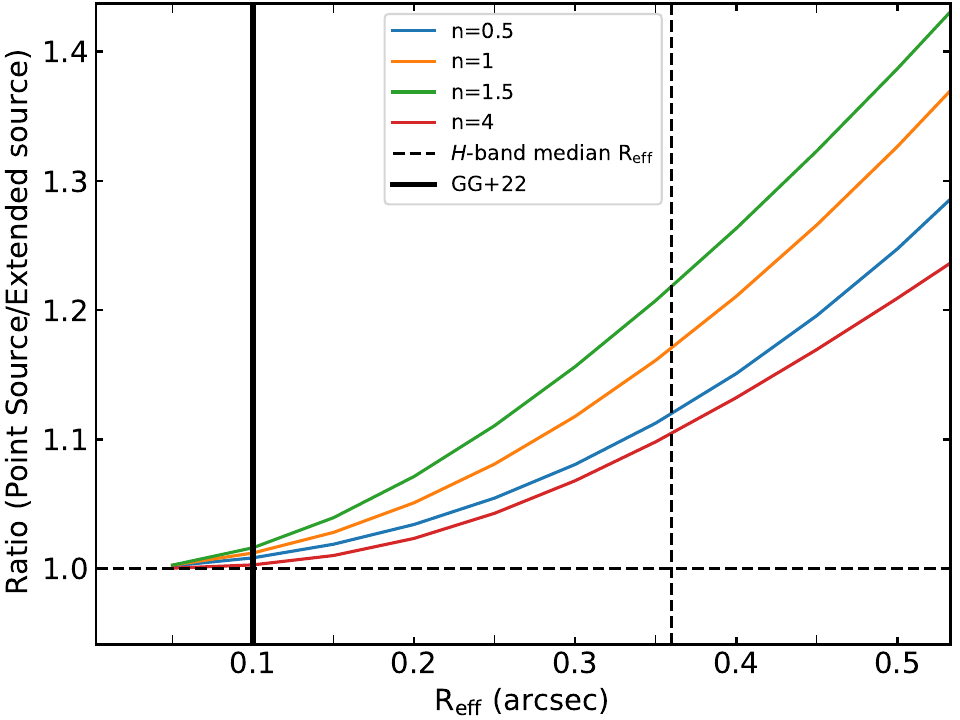}
\caption{Flux ratio between a point source emission and a modeled galaxy profile versus the effective radius. The blue line shows the profile for a S\'ersic index, n = 0.5, the orange one for n = 1, the green one for n = 1.5, and the red one for n=4. 
The vertical solid line shows the typical size of the dust component according to \citet{carlos2022a} for a $z=1.9$, log M$_\star$/M$_\odot\sim$~10.5 galaxy. The vertical dashed line shows the typical size according to the median S\'ersic parameters extracted from G13 for the M$_\star \geq 10^{10}\;$M$_\odot$ galaxies in our sample.}
\label{fig:sizes}
\end{figure}

According to GG22, the effective radius R$_{\mathrm{eff}}$ (the radius that contains half of the total light) of the dust component of a source with $z=1.9$ and M$_\star=10^{10.5}\;$M$_\odot$ is 0$\farcs$10.
At HST $H$-band resolution our galaxies are fitted by a S\'ersic profile characterized by a median S\'ersic index n = 1.36 and a median effective radius R$_{\mathrm{eff}}=0\farcs36$. In Fig.~\ref{fig:sizes}, we show the flux correction factor one should take for our measurements (i.e., at 10$^{10-11}$~M$_\odot$) versus the R$_{\mathrm{eff}}$. Focusing on the size estimation provided by GG22, the flux correction associated with our measurements is negligible. According to the size of the stellar component, for a S\'ersic index n = 1.0-1.5, this correction ranges from 1.17-1.22. If the size of the dust component resembled that of the stellar component, this $\sim$20\% correction would translate to 0.08~dex larger M$_{\textrm{gas}}$ than those reported in Table~\ref{tab:properties2} and Fig.~\ref{fig:fgas}. 

\section{Gas reservoirs} 
\label{sec:gas_prop}

\subsection{Observed evolution of the gas reservoir of our sample}
\label{sec:fgas}

We calculate the gas content of our sample following two approaches. The first is based on the computation of a $\delta_{\mathrm{GDR}}$ using a mass metallicity relation (MZR), and the second on the RJ dust continuum emission (see Sec.~\ref{sec:intro}). 

For the first one, we produce synthetic spectra of the dust emission of our galaxies, according to their median redshift and $\Delta$MS, using the \citet{Schreiber2018} IR SED template library\footnote{http://cschreib.github.io/s17-irlib/}. This library contains 300 templates, divided into two classes: 150 dust continuum templates due to the effect of big dust grains, and 150 templates that include the MIR features due to polycyclic aromatic hydrocarbon molecules (PAHs). These templates, which can be co-added, correspond to the luminosity that is emitted by a dust cloud with a mass equal to 1~M$_\odot$. After scaling each template to the measured flux density of the stacked galaxy at 1.1~mm, we obtain the LIR by integrating the rest-frame template flux between 8 and 1000$\mu$m. This luminosity is then translated to M$_\mathrm{dust}$ by multiplying the intrinsic M$_\mathrm{dust}$/LIR of the template by the LIR that corresponds to the measured flux density. \citet{Schreiber2018} models use different dust grain composition and emissivity yielding lower M$_\mathrm{dust}$ by a factor of two on average when compared to the more widely used \citet{DraineLi2007} models. Therefore, in order to have comparable M$_\mathrm{dust}$ with the literature studies and prescriptions needed to convert them into M$_{\mathrm{gas}}$, we re-scale the results based on the \citet{Schreiber2018} by an appropriate factor at each source redshift (Leroy et al. in prep). M$_{\mathrm{gas}}$ is then obtained through the dust emission using the $\delta_{\mathrm{GDR}}$-Z relation derived by \citet{Magdis2012}, assuming the MZR from \citet{Genzel2015}, using the median M$_\star$ and $z$ of the corresponding bin. The M$_{\mathrm{gas}}$ that we get using this approach corresponds to the total gas budget of the galaxies, including the molecular and atomic phases. As explained in \citet{Carlos2022b} and references therein, the molecular gas dominates over the atomic one within the physical scales probed by the dust continuum observations at this wavelength. It is worth noting that this statement has been tested within the angular scales probed by dust continuum observations, but the HI may dominate at larger scales \citep{Chowdhury2020}.

Let us note that this approach assumes that the emissivity index ($\beta$) adopted in the \citet{Schreiber2018} templates ($\sim$1.5, the average value for local dwarf galaxies; \citealt{Lyu2016}) is accurate for our galaxies since we do not have FIR data to better constrain this parameter. Leroy et al. (in prep) perform stacking using ALMA and $Herschel$ data to obtain the SED of typical MS galaxies. They obtain this $\beta$ by SED fitting, getting values that are compatible with the $\beta=1.5$ assumed in the \citet{Schreiber2018} models.
In \citet{Shivaei2022}, they use stacks of \textit{Spitzer}, $Herschel$, and ALMA photometry to examine the IR SED of high-z subsolar metallicity ($\sim$0.5~Z$\odot$) luminous IR galaxies (LIRGs), adopting $\beta=1.5$ for their sample. In this paper, they also discuss other possible values of this parameter, but still, they perform the analysis using $\beta=1.5$.

For the second approach, we follow S16, using the corrected version of equation 16 from that paper. In this paper, they affirm that the luminosity-to-mass ratio at 850~$\mathrm{\mu}$m is relatively constant under a wide range of conditions in normal star-forming and starburst galaxies, at low and high redshifts. We can thus use the measurements of the RJ flux density, derive the luminosity, and estimate M$_{\mathrm{gas}}$. They note that this approach is equivalent to a constant $\delta_{\mathrm{GDR}}$ for high stellar mass galaxies. This approach is justified if the variation of the mass-weighted T$_{\mathrm{dust}}$ on galactic scales is small. This is true for galaxies in the vicinity of the MS, as reported in \citet{Magnelli2014}, that uses stacked Herschel data up to $z\sim2$. However, higher and a wider range of temperatures are observed in systems further away from the MS (see \citealt{Clements2010}, \citealt{Cochrane2022}). The fact that the mass-weighted T$_{\mathrm{dust}}$ keeps constant at these redshifts for MS galaxies is also supported by simulations. \citet{Liang2019}, using the high-resolution cosmological simulations from the Feedback in Realistic Environments (\texttt{FIRE}) project, report that the mass-weighted T$_{\mathrm{dust}}$ does not strongly evolve with redshift over $z = 2-6$ at fixed IR luminosity. At a fixed redshift, it is however tightly correlated with LIR, hence sources with very high LIR, normally starburst objects, show higher mass-weighted T$_{\mathrm{dust}}$ than 25~K. We do not expect such high LIR values for our galaxies.

It is also important to take into account that the RJ tail methods can be safely applied if $\lambda_{\mathrm{rest}}>>$~hc/(k$_\mathrm{B}$T$_{\mathrm{dust}}$), where k$_\mathrm{B}$ is the Boltzmann constant and T$_{\mathrm{dust}}$ refers to the mass-weighted T$_{\mathrm{dust}}$. For a mass-weighted T$_{\mathrm{dust}}$ of 25 K, this requires $\lambda_{\mathrm{rest}}>>580$~$\mu$m. At higher redshifts and for lower mass-weighted T$_{\mathrm{dust}}$, the rest-frame wavelength probed by ALMA band-6 observations is far from the RJ regime \citep{Cochrane2022}. Results at $z\sim3$ should therefore be interpreted with caution.

S16 insist that the calibration samples that they use are intentionally restricted to objects with high stellar mass (M$_\star> 5 \times 10^{10}~$M$_\odot$), hence not probing lower metallicity systems. As a consequence, we only use this approach for the calculation of the M$_{\mathrm{gas}}$ in the high-mass bin. On the other hand, for the low- and intermediate-mass bins, the M$_{\mathrm{gas}}$ is only computed following the $\delta_{\mathrm{GDR}}$ method.
We discuss the effect of S16 and other prescriptions in the calculation of the gas content in lower mass galaxies in Sec.~\ref{sec:discussion}. 

An offset between both approaches, S16 and $\delta_{\mathrm{GDR}}$, is expected though at high stellar masses, as reported in \citet{Carlos2022b}, where they find a median relative difference (M$_{\mathrm{gas}}^{\mathrm{RJ}}$-M$_{\mathrm{gas}}^{\delta_{\mathrm{GDR}}}$)/M$_{\mathrm{gas}}^{\delta_{\mathrm{GDR}}}$=$~0.23\pm0.84$ between both measurements.

The errors are estimated through 10,000 Monte-Carlo simulations perturbing the photometry randomly within the total uncertainties and assuming an error of 0.20~dex for the metallicity \citep{Magdis2012}. All these values are included in Table~\ref{tab:properties2}. As done previously for the fluxes, we also computed the error due to the underlying distribution for the median log M$_\star$, $z$, and $\Delta$MS using bootstrapping. When propagating these uncertainties (of the order of the hundredth) to derive the error in the metallicity we obtain values lower than 0.20~dex, and hence lower uncertainties for the metallicity-derived parameters. As a consequence, we decided to keep the 0.2~dex value for the metallicity uncertainty, which is more conservative.

\begin{figure*}[ht]
    \centering
    \includegraphics[width=18.6cm, height=5.3cm]{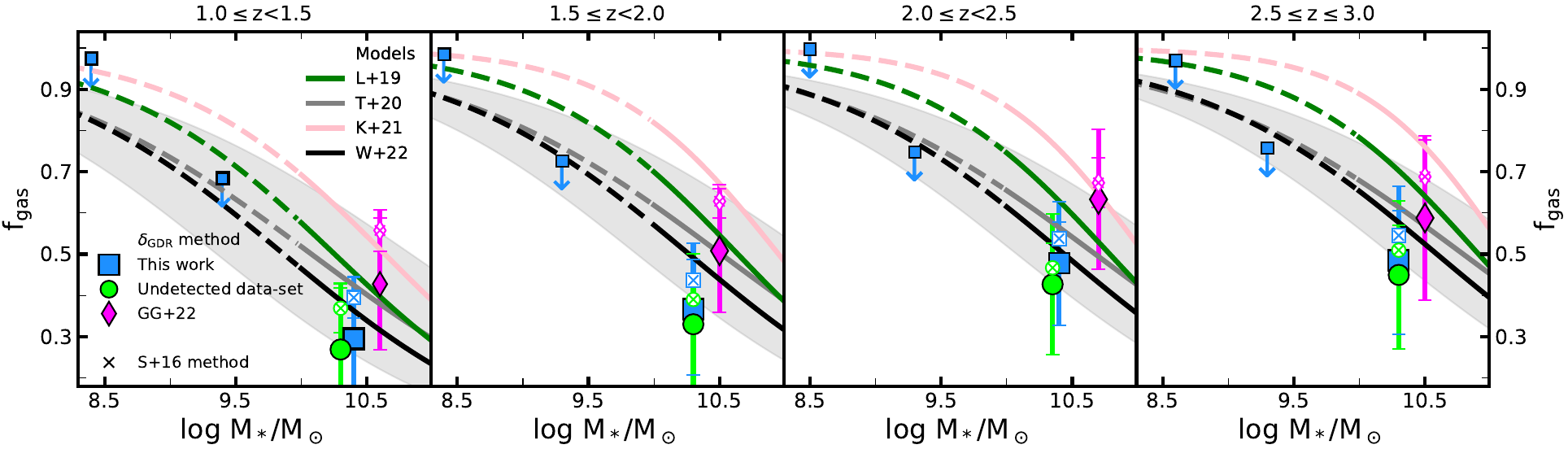}
    
    \caption{Gas fractions versus stellar masses derived for each redshift range. Squares represent the gas fractions obtained for our sample, circles show the gas fractions derived for the undetected data-set. Diamonds represent the gas fractions calculated for the GG22 galaxies only. The differences between samples highlight the effect of the exclusion and inclusion of individually detected galaxies in the gas fractions. Gas fractions from the colored filled markers are computed following the gas-to-dust ratio prescription whereas colored empty markers with a cross within represent the gas fractions as calculated using the S16 method (only for the high-mass bin, upper limits are only computed following the gas-to-dust ratio method). Both values are included in Table~\ref{tab:properties2}. 3$\sigma$ upper limits at lower stellar masses are shown with smaller squares and vertical arrows. Uncertainties are also included for the measurements. We also show the \citet{Liu2019b} (L19), \citet{Tacconi2020} (T20), \citet{Kokorev2021} (K21), and \citet{Wang2022} (W22) scaling relations in green, gray, pink, and black, respectively. For each line, there is a dashed and a solid part. The solid part represents the mass range for which these relations are calibrated, whereas the dashed one shows an extrapolation of these relations for lower stellar masses. For the T20 relation, we also include the uncertainty as a shaded gray region. The scaling relations are computed using the median redshift of each bin and a $\Delta$MS = 0. The distance from each of the points to these scaling relations is re-scaled to their corresponding redshifts and $\Delta$MS using T20.}
    \label{fig:fgas}
\end{figure*}

To fully understand and check the consistency of our measurements, we include two additional calculations of the gas fractions for the high-mass bin: one for the undetected data-set, and another considering only the GG22 galaxies. This is to both compare the measurement of the gas fractions that we obtain when considering the full mass-complete sample with the results excluding individually detected sources and to check what is the contribution of these bright individually-detected galaxies in the stacked measurements. All these values are also included in Table~\ref{tab:properties2}. 

In Fig.~\ref{fig:fgas}, we show the evolution with M$_\star$ of the gas fractions derived for each redshift bin. As mentioned in Sec.~\ref{sec:stacking}, we provide 3$\sigma$ upper limits for the low- and intermediate-mass bins and measurements for the high-mass bin. For the low-mass bin we get f$_{\mathrm{gas}}<0.97-0.98$ along $1\leq z\leq3$. These numbers are f$_{\mathrm{gas}}<0.69-0.77$ for the intermediate-mass bin. For the high-mass bin, focusing first on the $\delta_{\mathrm{GDR}}$ results, we obtain f$_{\mathrm{gas}}=0.32-0.48$. Looking at the two additional cases, we check that removing the GG22 galaxies and their neighbors from our sample (i.e., considering only the undetected data-set) slightly drops the measurements (f$_{\mathrm{gas}}=0.30-0.45$). The number of detected objects is much lower compared to the contribution of the undetected galaxies (see Table~\ref{tab:properties2}), which dominate the emission from the stack. If we only consider the detected galaxies from GG22 and stack them following the procedure described along Sec.~\ref{sec:stacking}, we get, as expected, much higher gas fractions (f$_{\mathrm{gas}}=0.46-0.66$). Taking the individual gas fractions of the GG22 objects, provided in \citet{Carlos2022b}, and calculating the mean for each redshift bin, we get values of f$_{\mathrm{gas}}=0.53-0.69$. If we turn to the gas fractions as derived using the S16 relation, we see that the latter provides higher values of the gas fractions but both results are compatible within the uncertainties.

If we compare the values of f$_{\mathrm{gas}}$ in terms of the kind of stacking method chosen, we see that mean stacking (see Appendix~\ref{app:mean_stack}) provides slightly higher values of the gas content at $10^{10-11}$~M$_\odot$ (f$_{\mathrm{gas}}=0.34-0.54$) getting closer to the W22 and T20 relations. Overall, this variation is compatible within the uncertainties derived for f$_{\mathrm{gas}}$.

\subsection{Scaling relations framework and comparison with our sample}
\label{sec:scaling}

Based on galaxy samples such as the ones described in Sec.~\ref{sec:properties}, several works provide us with different scaling relations that allow us to obtain the gas content of galaxies given the redshift, the $\Delta$MS, and the M$_\star$. Some of these are \citet{Liu2019b} (L19), T20, \citet{Kokorev2021} (K21), and \citet{Wang2022} (W22).

The L19 relation is based on the A$^3$COSMOS project, already introduced in Sec.~\ref{sec:properties}, together with $\sim$1,000 CO-observed galaxies at $0<z<4$ (75\% of them at $z<0.1$). The galaxies from the A$^3$COSMOS project probe the log M$_\star/$M$_\odot\sim11-12$ MS. Complementary sources (most of them belonging to \citealt{Tacconi2018} and \citealt{Kaasinen2019}) sample the log M$_\star/$M$_\odot\sim10-11$ MS at $z>1$. For $z<0.03$, the complementary sample also covers the log M$_\star/$M$_\odot\sim9-10$ MS, but they insist that the metallicity-dependent CO-to-H$_2$ conversion factor $\alpha_{\mathrm{CO}}$ might be more uncertain, and so the estimated M$_{\mathrm{gas}}$.

T20 is based on individually detected objects plus stacks of fainter galaxies, as pointed out in Sec.~\ref{sec:properties}. This relation is an expansion of the results obtained in \citet{Tacconi2018}.

K21 uses $\sim$5,000 SFGs at $z<4.5$, drawn from the super-deblended catalogs introduced in Sec.~\ref{sec:properties}. The median redshift of the sample K21 is based on is $z\sim0.90$, with a median M$_\star\sim$~4.07$\times$10$^{10}$~M$_\odot$. The low-mass and low-redshift part of their sample is restricted to galaxies that lie above the MS. Nevertheless, 69\% of the galaxies qualify as MS, 26\% are classified as starbursts and 5\% qualify as passive galaxies. 

W22 is based on the COSMOS2015 galaxy catalog (in Sec.~\ref{sec:properties} we referred to COSMOS2020, which is an updated version of COSMOS2015). They select star-forming MS galaxies with ALMA band 6 or 7 coverage in the A$^3$COSMOS database and well within the ALMA primary beam, obtaining a final sample of 3,037 sources. They stack galaxies, binning in redshift and M$_\star$, in the $uv$ domain, covering the M$_\star$ range 10$^{10-12}$~M$_\odot$. They do not select galaxies according to a certain SNR threshold at (sub-)~millimeter wavelengths, so the sample includes both, detected and undetected ALMA sources. 

In Fig.~\ref{fig:fgas}, the values of the scaling relations there represented are derived using the median redshift of the bin and a $\Delta$MS = 0. 

If we compare our results with the previous scaling relations, for the high-mass bin, we see that the measurements for our sample are more compatible with the W22 scaling relation than with any of the latter relations. These measurements are also within the uncertainties defined by T20 but below L19 and K21. In the case of the intermediate-mass bin, the upper limits lie on the level established by the W22 and T20 extrapolations, and well below L19 and K21. In the low-mass bin, upper limits are poorly constraining and lie above most of the scaling relations.

\section{Discussion} \label{sec:discussion}

Pushing the limit to bluer, less dusty, more MS-like, and more mass-complete samples yields lower levels of gas than those prescribed by literature scaling relations based on redder and less complete data-sets. The super-deblended catalogs and the A$^3$COSMOS samples are highly dust-obscured, showing red optical colors, and are more star-forming than our galaxies. As a consequence, the L19 and K21 scaling relations yield f$_{\mathrm{gas}}-$M$_\star$ relations with a higher normalization. Going to mass-complete samples, like the one used in the W22 relation, leads to the inclusion of blue objects with low obscurations and SFRs compatible with being right on top of the MS. As a result, the W22 relation exhibits a lower normalization, better matching our results. T20 lies between the two regimes, presumably because it is based on a combination of individually detected red, dust-obscured objects, complemented by stacks of bluer and fainter galaxies.

Regarding our results for the high-mass bin, our low values of f$_{\mathrm{gas}}$ could still be contaminated to a certain extent by the presence of post-starburst galaxies in their way to quiescence. These galaxies can have passed the $UVJ$ screening due to their blue $U-V$ color and can be pulling the f$_{\mathrm{gas}}$ to lower values. It is true, however, that this effect gains importance at $z>3$ (\citealt{Deugenio2020}, \citealt{Forrest2020}, \citealt{Valentino2020}), out of the redshift range considered in this work. In \citet{Antwi-Danso2023}, they test the performance of the $UVJ$ diagram selecting quiescent galaxies, including post-starbursts. According to their results based on the \texttt{Prospector-$\alpha$} SED modeling, the $UVJ$ selection reaches the $\sim$90\% completeness at $z\leq4$. They define this completeness as the number of quiescent galaxies that are selected divided by the total number of quiescent galaxies in the sample, with quiescence being defined as a specific SFR below the threshold of the green valley.

To quantify the effect of the possible contamination due to these post-starburst objects, we produce mock sources, based on sky positions where no galaxies have been cataloged, and introduce them in the stack, checking their imprint on the resulting f$_{\mathrm{gas}}$. Considering the $UVJ$ selection to be 90\% complete at these redshifts, we see that after introducing these mock sources we obtain between 5\%-7\% less f$_{\mathrm{gas}}$. This difference is smaller than the uncertainty we derive for this parameter.

Additionally, the fact that we are comparing our values of f$_{\mathrm{gas}}$, obtained using a certain method, with the results provided by scaling relations whose measurements of f$_{\mathrm{gas}}$ come from different conversion prescriptions, might be another source of discrepancy.

The L19 and W22 scaling relations rely on the RJ-tail continuum method of \citet{Hughes2017}. Using this prescription to compute the f$_{\mathrm{gas}}$ of our sample, adopting $\alpha_{\textrm{CO}}$=6.5~(K km s$^{-1}$ pc$^2)^{-1}$, we get 7\% larger values, similarly to what we get using S16. T20 use the \citet{Leroy2011} $\delta_{\mathrm{GDR}}$ together with the \citet{Genzel2015} MZR. Using this prescription, we obtain 3\% larger values of f$_{\mathrm{gas}}$. K21 is based on the \citet{Magdis2012} $\delta_{\mathrm{GDR}}$ prescription together with the \citet{Mannucci2010} fundamental metallicity relation (FMR) calibrated for \citet{KewleyDopita2002} that they convert to the \citet{PettiniPagel2004} (PP04 N2) scale following \citet{Kewley2008}. Using this method, we get 5\% larger values of f$_{\mathrm{gas}}$.

The discrepancy between some of the scaling relations and our data is therefore not a consequence of the methodology or other factors that might be artificially pulling down our values, but simply the end result of considering a mass-complete sample that includes bluer less-dusty objects in comparison with other samples progressively more and more biased to redder dustier galaxies.

Concerning our findings for the intermediate-mass bin, it is important to take into account that at low M$_\star$ the link between metallicity and the $\alpha_{\mathrm{CO}}$ or the $\delta_{\mathrm{GDR}}$ is still not well constrained and can lead to a bad estimation of the gas content. Up to date, there is very little information about the gas content of galaxies with $\sim10^9$~M$_\odot$ at high redshifts. Most efforts so far focused on galaxies at $z\sim0$ (e.g., \citealt{Jiang2015}, \citealt{Cao2017}, \citealt{Saintonge2017}, \citealt{Madden2020}, \citealt{Leroy2021}). According to T20 and references therein, it is hard or impossible to detect low-mass galaxies with substantially subsolar metallicity and to determine their gas content quantitatively. They suggest that there might be an interstellar medium component that might be missed or overlooked with the current techniques, such as gas/dust at very low temperatures. Deeper observations would be required to provide a better constraint on the f$_{\mathrm{gas}}$ of these systems.

We also test the effect of using different prescriptions to compute the f$_{\mathrm{gas}}$ in the intermediate-mass bin. Using RJ continuum methods such as S16 or \citet{Hughes2017} yields $\sim10$\% lower values than the ones we report. Discrepancies are expected since these methods are calibrated for more massive galaxies. S16 relies on a sample of 0.2$-$4$\times10^{11}$~M$_\odot$ galaxies whereas the \citet{Hughes2017} sample comprises M$_\star$ ranging from $6-11\times 10^{10}$~M$_\odot$. On the other hand, the \citet{Leroy2011} prescription provides values which are compatible with our results (they differ in less than 1\%), whereas the use of the \citet{Magdis2012} prescription using the \citet{Mannucci2010} FMR yields similar f$_{\mathrm{gas}}$ at $z<2$ but starts differing at higher redshifts, where this approach reports 8\% less f$_{\mathrm{gas}}$. This difference is compatible with the uncertainties but still could reflect that the metallicity of low-mass galaxies at $z>2$ deviates from that observed for local galaxies, contrary to what is seen in higher-mass systems, whose metallicity does not evolve with redshift until $z>2.5$ \citep{Mannucci2010}. This highlights the need to re-calibrate these relations for less massive objects compatible with being MS galaxies. In most cases, the low-mass sample of this kind of studies are mainly made up of galaxies showing very high SFRs.

\section{Summary and conclusions} \label{sec:conclusions}

Taking advantage of the CANDELS mass-complete catalog performed in GOODS-S \citep{Guo2013}, we are able to explore the gas content of galaxies in ALMA, using Band-6 observations at 1.1~mm \citep{carlos2022a}. Our sample is composed of 5,530 star forming blue ($<b-i>\sim0.12$~mag, $<i-H>\sim$0.81~mag) galaxies at $1.0\leq z\leq3.0$, located in the main sequence. It allows us to explore the gas content of $10^{10-11}$~M$_\odot$ star-forming galaxies regardless of their emission at (sub-)millimeter wavelengths. Additionally, and thanks to the stellar mass coverage and completeness of the sample, we can provide an upper limit of the gas content of lower mass galaxies at $\sim10^{9-10}$~M$_\odot$.
We report measurements at $10^{10-11}$~M$_\odot$ and 3$\sigma$ upper limits for the gas fraction at $10^{8-10}$~M$_\odot$.

At $10^{10-11}$~M$_\odot$, we are tracing lower gas fractions, f$_{\mathrm{gas}}=0.32-0.48$, than those derived from other scaling relations that use samples of redder and dustier objects on average, being biased towards individually-detected sources at (sub-)millimeter wavelengths, more subject to higher attenuations and also more star-forming than our galaxies. Relations based on more general mass-complete samples show more compatible values to the ones we report.

At $10^{8-9}$~M$_\odot$, the values we retrieve lie well above the scaling relations extrapolation, whereas at $10^{9-10}$~M$_\odot$ the upper limits, ranging from 0.69 to 0.77, are located well within the region defined by the \citet{Wang2022} and \citet{Tacconi2020} scaling relations. The position of the upper limits at these intermediate masses supports the idea that the extrapolation derived from these scaling relations is representative of the upper bound of the underlying f$_{\mathrm{gas}}-$M$_\star$ relation as traced by the bulk of star-forming galaxies.

\begin{sidewaystable*}[ht]
\setlength{\tabcolsep}{2.5pt} 
\renewcommand{\arraystretch}{1.2}
\centering
\caption{Summary of the physical properties derived in this work for our sample, the undetected data-set, and GG22. The T$_\mathrm{dust}$ are those corresponding to the \citet{Schreiber2018} templates derived for the galaxies. The subindex GDR denotes that the quantity has been calculated using the gas-to-dust ratio approach. The subindex S16 means that the quantity is derived following \citet{Scoville2016}. The lack of uncertainty in the gas fractions, gas, and dust masses, and luminosities denotes an upper limit (3$\sigma$ level). Let us recall that for the undetected data-set we also remove the G13 sources in the neighborhood of the GG22 galaxies. We have 3 of these neighbor objects at $1<z<1.5$ and another 3 at $1.5<z<2.0$ in the high-mass bin; 2 at $2.0<z<2.5$, and none at $2.5<z<3.0$.}
\label{tab:properties2}
\begin{tabular}{l|llllllllllll}
z bin&log M$_\star$ bin&N$_{\textrm{obj}}$&z$_{\mathrm{median}}$&log M$_{\mathrm{*, median}}$&$\Delta$MS$_{\mathrm{median}}$&log LIR& T$_{\mathrm{dust}}$&log M$_{\mathrm{dust}}$&log M$_{\mathrm{gas,GDR}}$&log M$_{\mathrm{gas,S16}}$& f$_{\mathrm{gas,GDR}}$&f$_{\mathrm{gas,S16}}$\\ &(M$_\odot$)& &&(M$_\odot$)&(dex) &(L$_\odot$)&(K)&(M$_\odot$)&(M$_\odot$)&(M$_\odot$)& & \\
\hline\hline
&&&&&\textbf{Whole sample} \\
\hline
        &$8\leq$~log M$_\star<$~9 &910&1.3&8.41 & $-0.02$ &10.26&29.36&7.15 & 10.00&&$<$~0.98&\\
      1.0~$\leq z<$~1.5&$9\leq$~log M$_\star<$~10 &299& 1.2& 9.37& 0.00 &10.43 &29.36&7.35 &9.72&&$<$~0.69&\\
     &$10\leq$~log M$_\star<$~11 &92& 1.2 & 10.35&0.07&11.18$\pm$0.10&30.04&7.97$\pm$0.10& 10.02$\pm$0.30&10.20$\pm$0.10&0.32$\pm$0.15&0.42$\pm$0.05 \\
\hline
       &$8\leq$~log M$_\star<$~9 &1423&1.7 &8.41 & $-0.08$ &10.21&30.84 &6.97 &9.97&&$<$~0.97&\\
      1.5~$\leq z<$~2.0&$9\leq$~log M$_\star<$~10&353 & 1.8& 9.31& 0.13 &10.66&33.16&7.28 &9.80&&$<$~0.75& \\
     &$10\leq$~log M$_\star<$~11 &96& 1.8 & 10.30&0.17&11.44$\pm$0.09&33.49&8.00$\pm$0.09 & 10.14$\pm$0.29&10.27$\pm$0.09&0.41$\pm$0.16&0.48$\pm$0.05 \\
\hline
     &$8\leq$~log M$_\star<$~9 &881&2.3 &8.50& $-0.15$ &10.38 &32.59&7.04 &10.12&&$<$~0.98&\\
      2.0~$\leq z<$~2.5&$9\leq$log M$_\star<$10 &274& 2.3& 9.31& 0.04 &10.74 &34.70& 7.17 &9.80&&$<$~0.76&\\
     &$10\leq$~log M$_\star<$~11&54 & 2.2 & 10.38&$-0.03$&11.61$\pm$0.07&33.59 & 8.16$\pm$0.07 & 10.34$\pm$0.27&10.43$\pm$0.07&0.47$\pm$0.15&0.53$\pm$0.04\\
\hline
     &$8\leq$~log M$_\star<$~9 &751&2.7 &8.56& 0.00 &10.58 &36.21&6.90 &10.03&&$<$~0.97&\\
      2.5~$\leq z\leq$~3.0&$9\leq$~log M$_\star<$~10 &340& 2.8& 9.29& 0.06 &10.84 &37.00& 7.12 &9.82&&$<$~0.77&\\
     &$10\leq$~log M$_\star<$~11 &57& 2.7 & 10.32&$-0.02$&11.71$\pm$0.09&36.03 & 8.04$\pm$0.11 & 10.29$\pm$0.31&10.39$\pm$0.11&0.48$\pm$0.18&0.54$\pm$0.06\\
\hline
&&&&&\textbf{Undetected data-set} \\
\hline
     1.0~$\leq z<$~1.5&$10\leq$~log M$_\star<$~11 &85& 1.2 & 10.34&0.06&11.12$\pm$0.11&29.93&7.91$\pm$0.11& 9.97$\pm$0.31&10.15$\pm$0.11&0.29$\pm$0.15&0.39$\pm$0.06 \\
      1.5~$\leq z<$~2.0&$10\leq$~log M$_\star<$~11&84 & 1.8 & 10.29&0.12&11.28$\pm$0.12&33.01&7.90$\pm$0.12& 10.05$\pm$0.31&10.14$\pm$0.12&0.36$\pm$0.17&0.42$\pm$0.06 \\
     2.0~$<z\leq$~2.5&$10\leq$~log M$_\star<$~11&46 & 2.2 & 10.37&$-0.03$&11.49$\pm$0.10&33.64 & 8.04$\pm$0.10 & 10.22$\pm$0.30&10.30$\pm$0.10&0.42$\pm$0.17&0.46$\pm$0.06\\
     2.5~$\leq z\leq$~3.0&$10\leq$~log M$_\star<$~11&47 & 2.7 & 10.27&0.00&11.60$\pm$0.11&36.17 & 7.92$\pm$0.11 & 10.19$\pm$0.31&10.29$\pm$0.11&0.45$\pm$0.18&0.51$\pm$0.06\\
\hline
&&&&&\textbf{G22} \\
\hline
      1.0~$\leq z<$~1.5&$10\leq$~log M$_\star<$~11 &4& 1.2 & 10.63&0.10&11.77$\pm$0.09&30.33&8.56$\pm$0.09& 10.55$\pm$0.29&10.79$\pm$0.09&0.46$\pm$0.16&0.59$\pm$0.05 \\
      1.5~$\leq z<$~2.0&$10\leq$~log M$_\star<$~11 & 9&1.9 & 10.53&0.34&12.31$\pm$0.08&36.02&8.65$\pm$0.08 & 10.74$\pm$0.28&10.99$\pm$0.08&0.62$\pm$0.15&0.74$\pm$0.04 \\
      2.0~$< z\leq$~2.5&$10\leq$~log M$_\star<$~11&6 & 2.2 & 10.65&0.09&12.40$\pm$0.12&34.77& 8.84$\pm$0.12 & 10.94$\pm$0.32&11.02$\pm$0.12&0.66$\pm$0.17&0.70$\pm$0.06\\
       2.5~$\leq z\leq$~3.0&$10\leq$~log M$_\star<$~11 &10& 2.8 & 10.49&0.11&12.26$\pm$0.18&37.43& 8.51$\pm$0.18 & 10.71$\pm$0.37&10.89$\pm$0.18&0.62$\pm$0.20&0.72$\pm$0.09\\
\hline
\end{tabular}
\end{sidewaystable*}

\begin{acknowledgements}
      RMM acknowledges support from  Spanish  Ministerio de Ciencia e Innovaci\'on MCIN/AEI/10.13039/501100011033 through grant PGC2018-093499-B-I00, as well as from MDM-2017-0737 Unidad de Excelencia “Maria de Maeztu” - Centro de Astrobiología (CAB), CSIC-INTA, “ERDF A way of making Europe”, and INTA SHARDS$^{JWST}$ project through the PRE-SHARDSJWST/2020 PhD fellowship. CGG acknowledges support from CNES. PGP-G acknowledges support from grants PGC2018-093499-B-I00 and PID2022-139567NB-I00 funded by Spanish Ministerio de Ciencia e Innovación MCIN/AEI/10.13039/501100011033, FEDER, UE. PSB acknowledges support from Spanish Ministerio de Ciencia e Innovaci\'on MCIN/AEI/10.13039/501100011033 through the research projects with references PID2019-107427GB-C31 and PID2022-138855NB-C31. MF acknowledges NSF grant AST-2009577 and NASA JWST GO Program 1727. GEM acknowledges the Villum Fonden research grant 13160 “Gas to stars, stars to dust: tracing star formation across cosmic time”, grant 37440, “The Hidden Cosmos”, and the Cosmic Dawn Center of Excellence funded by the Danish National Research Foundation under the grant No. 140. This work has made use of the Rainbow Cosmological Surveys Database, which is operated by the Centro de Astrobiología (CAB/INTA), partnered with the University of California Observatories at Santa Cruz (UCO/Lick, UCSC). We made use of the following ALMA data: ADS/JAO.ALMA\#2015.1.00543.S and ADS/JAO.ALMA\#2017.1.00755.S. ALMA is a partnership of ESO (representing its member states), NSF (USA), and NINS (Japan), together with NRC (Canada), MOST and ASIAA (Taiwan), and KASI (Republic of Korea), in cooperation with the Republic of Chile. The Joint ALMA Observatory is operated by ESO, AUI/NRAO and NAOJ.
\end{acknowledgements}

%
%

\bibliographystyle{aa} 
\bibliography{ALMA_AA} 

\begin{appendix}

\section{Mean stacking}
\label{app:mean_stack}

\begin{table*}[ht]
\setlength{\tabcolsep}{6.5pt} 
\renewcommand{\arraystretch}{1}
\centering
\caption{Analog of Table~\ref{tab:properties1} based on mean stack measurements including also the gas fractions for the different data-sets (last two columns). The subindex GDR denotes that the quantity has been calculated using the gas-to-dust ratio approach. The subindex S16 means that the quantity is derived following \citet{Scoville2016}.}
\label{tab:properties1_mean}
\begin{tabular}{lllllllll}
\hline\hline
\raggedleft
z bin&log M$_\star$ bin&N$_{\mathrm{obj}}$&Flux density &erFlux$_{\mathrm{stack}}$&erFlux$_{\mathrm{ind}}$&erFlux$_{\mathrm{total}}$&f$_{\mathrm{gas, GDR}}$&f$_{\mathrm{gas, S16}}$ \\&(M$_\odot$)& &($\mu$Jy)&($\mu$Jy)&($\mu$Jy)&($\mu$Jy)& &\\
\hline
& \textbf{Whole}& \textbf{sample} &&\\
\hline
        &$8\leq$~log M$_\star<$~9 &910 &$<$~16.57 & -&- &- & 0.97&-\\
      1.0~$\leq z<$~1.5&$9\leq$~log M$_\star<$~10 & 299& $<$~26.75& - &- &- &0.69 &-\\
     &$10\leq$~log M$_\star\leq$~11 & 92 & 146.69&19.88 &18.28 &27.01& 0.34$\pm$0.14&0.44$\pm$0.05\\
\hline
       &$8\leq$~log M$_\star<$~9 &1423 &$<$~11.91 & -  &-&-&0.97 &-\\
      1.5~$\leq z<$~2.0&$9\leq$log M$_\star<$10 & 353& $<$~26.24& -  &- &-&0.76 &-\\
     &$10\leq$~log M$_\star\leq$~11 & 96 & 187.63&18.49 &29.94&35.19&0.47$\pm$0.16&0.54$\pm$0.05\\
\hline
     &$8\leq$~log M$_\star<$~9 &881 &$<$~14.68 & -  &-&-&0.98 &-\\
    2.0~$<z<$~2.5&$9\leq$~log M$_\star<$~10 & 274& $<$~27.58& - &-&-&0.77&- \\
     &$10\leq$~log M$_\star\leq$~11 & 54 & 233.99&23.56 &42.40&48.51 &0.49$\pm$0.17&0.54$\pm$0.05\\
\hline
     &$8\leq$~log M$_\star<$~9 &751 &$<$~13.76 & -&- &-&0.96 &-\\
     2.5~$\leq z\leq$~3.0&$9\leq$log M$_\star<$10 & 340& $<$~25.16& - &-&-  &0.76 &-\\
     &$10\leq$~log M$_\star\leq$~11 & 57 & 269.35&24.75 &46.05 &52.28&0.54$\pm$0.16&0.60$\pm$0.05\\
\hline
& \textbf{Undetected}& \textbf{sources} &&\\
\hline
        &$8\leq$~log M$_\star<$~9 &890 &$<$~17.39 & - &- &-&0.97 &-\\
      1.0~$\leq z<$~1.5&$9\leq$~log M$_\star<$~10 & 285& $<$~26.52& -  &-&-& 0.69&- \\
     &$10\leq$~log M$_\star\leq$~11 & 85 & 121.38&21.28 &16.95&27.21& 0.30$\pm$0.14&0.40$\pm$0.05\\
\hline
       &$8\leq$~log M$_\star<$~9 &1387 &$<$~11.63 & - &- & -&0.97&-\\
      1.5~$\leq z<$~2.0&$9\leq$~log M$_\star<$~10 & 338& $<$~28.27& - &- &- &0.77 &-\\
     &$10\leq$~log M$_\star\leq$~11 & 84 & 109.58&19.75&17.29 &26.25 &0.36$\pm$0.16&0.41$\pm$0.06\\
\hline
     &$8\leq$~log M$_\star<$~9 &854 &$<$~14.31 & -  &-&-& 0.98&-\\
    2.0~$<z<$~2.5&$9\leq$~log M$_\star<$~10 & 265& $<$~29.75& -  &-&-& 0.77&-\\
     &$10\leq$~log M$_\star\leq$~11 & 46 & 154.37&26.16 &26.81 &37.46& 0.40$\pm$0.17&0.45$\pm$0.06\\
\hline
     &$8\leq$~log M$_\star<$~9 &732 &$<$~13.40 & - &-&-& 0.96&-\\
     2.5~$\leq z\leq$~3.0&$9\leq$~log M$_\star<$~10 & 325& $<$~25.12& - &- &- &0.78 &-\\
     &$10\leq$~log M$_\star\leq$~11 & 47 & 161.90&25.84  &27.32&37.60&0.45$\pm$0.17&0.51$\pm$0.06\\
\hline
\end{tabular}
\end{table*}

In table~\ref{tab:properties1_mean} we include the fluxes and f$_{\mathrm{gas}}$ that we obtain from mean stacking the galaxies.

\begin{figure*}
    \centering
   \includegraphics[width=16.1cm, height=23.5cm]{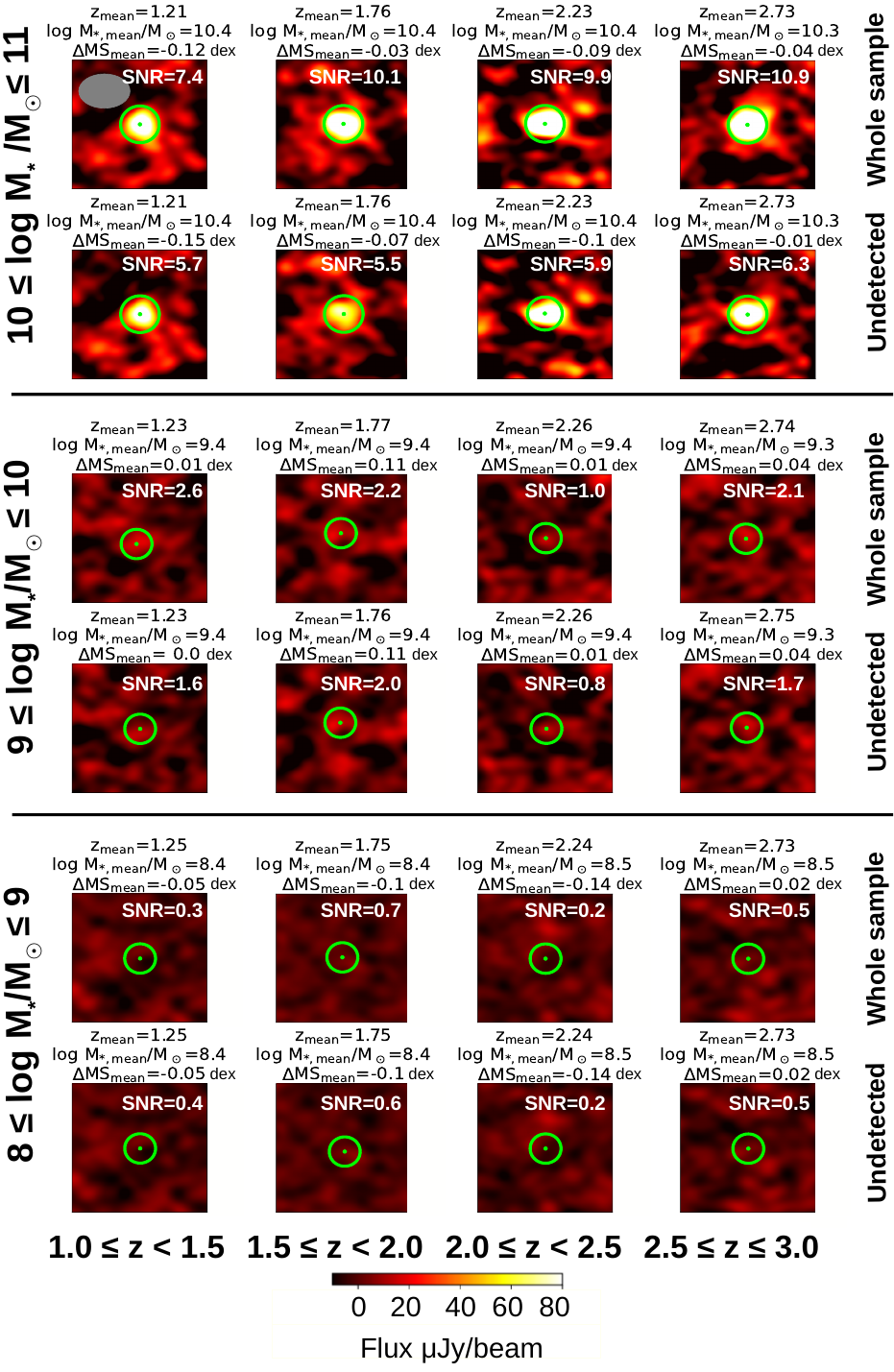}
    \caption{Cutouts of 7$\times$7 arcsec$^2$ of the ALMA low-resolution map showing the mean stacked galaxies in each redshift and mass bin (see Fig.~\ref{fig:mass_bins}) The flux densities and the corresponding uncertainties for each stacked galaxy are included in Table~\ref{tab:properties1_mean}.}
    \label{fig:mass_bins_mean}
\end{figure*}

\begin{figure*}[ht]
    \centering
    \includegraphics[width=18.6cm, height=5.3cm]{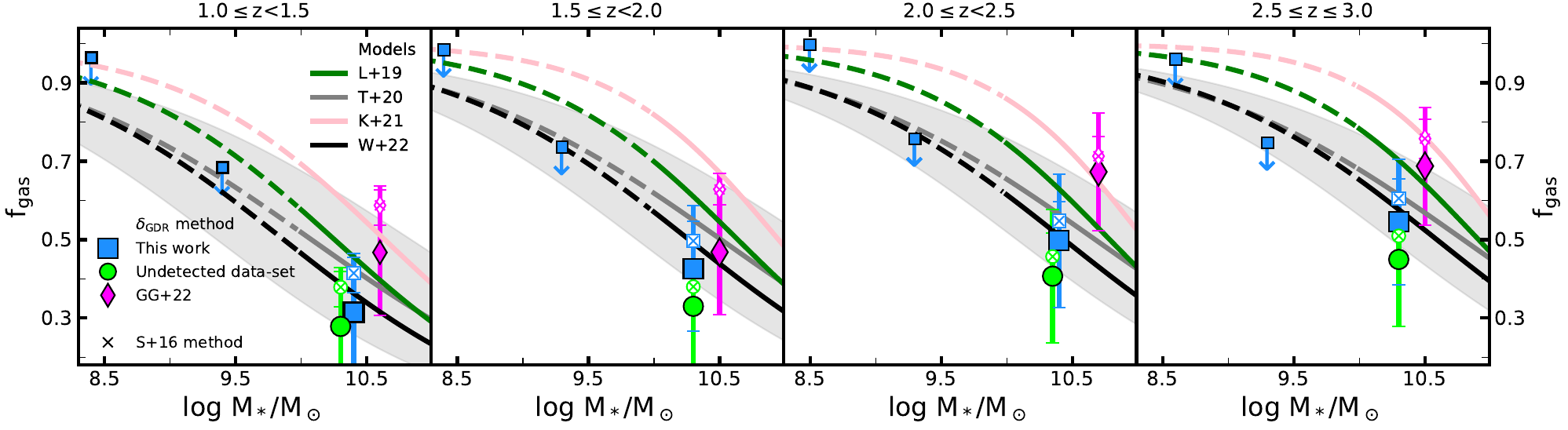}
    
    \caption{Gas fractions versus stellar masses derived for each redshift range, derived from mean stacking (also for the GG22 sample). See Fig.~\ref{fig:fgas} for a description of the markers and color codes here shown.}
    \label{fig:fgas_mean}
\end{figure*}

In Fig.~\ref{fig:mass_bins_mean} we include cutouts of the resulting stacked galaxies and Fig.~\ref{fig:fgas_mean} is an analog of Fig.~\ref{fig:fgas}, showing the results obtained from mean stacking.

\end{appendix}

\end{document}